\documentclass[11pt, a4paper]{article}
\usepackage{jcappub}

\def\mr{\mathrm}

\def\nm{\nonumber}

\def\pd{\partial}

\def\ol{\overline}
\def\mr{\mathrm}

\def\bs{\boldsymbol}
\def\nm{\nonumber}
\def\pd{\partial}

\def\pO{{\rm (O)}}
\def\pI{{\rm (I)}}
\def\pII{{\rm (II)}}
\def\pIIL{{\rm (II;L)}}
\def\pIIQ{{\rm (II;Q)}}

\newcommand{\dif}[2]{\frac{\mr{d} #1}{\mr{d} #2}}

\newcommand{\pdif}[2]{\frac{\pd #1}{\pd #2}}

\newcommand{\tetrad}[2]{{e^{(#1)}}_{#2}}
\newcommand{\itetrad}[2]{{e_{(#1)}}^{#2}}
\newcommand{\spin}[3]{{{\tilde{\omega}{}_{(#1)}}^{(#2)}}{}_{(#3)}}
\newcommand{\spinl}[3]{\tilde{\omega}{}_{(#1)(#2)(#3)}}
\newcommand{\ctetrad}[2]{{\tilde{e}{}^{(#1)}}_{#2}}
\newcommand{\citetrad}[2]{{\tilde{e}{}_{(#1)}}^{#2}}

\begin{document}

\title{Geodesic {\it curve}-of-sight formulae for the cosmic microwave background: 
a unified treatment of redshift, time delay, and lensing}
\author{Ryo Saito$^{1}$, Atsushi Naruko$^{2}$, Takashi Hiramatsu$^{3}$, and Misao Sasaki$^{3}$}
\affiliation{
$^1$ APC, (CNRS-Universit\'{e} Paris 7), 10 rue Alice Domon et L\'{e}onie Duquet, 75205 Paris, France,\\
$^2$ Department of Physics, Tokyo Institute of Technology, Tokyo 152-8551, Japan\\
$^3$ Yukawa Institute for Theoretical Physics, Kyoto University, Kyoto 606-8502, Japan\\
}

\abstract{
In this paper, 
we introduce a new approach to a treatment of the gravitational effects (redshift, time delay and lensing) on the observed cosmic microwave background (CMB) anisotropies based on the Boltzmann equation. 
From the Liouville's theorem in curved spacetime, 
the intensity of photons is conserved along a photon geodesic when non-gravitational scatterings are absent.
Motivated by this fact, 
we derive a second-order line-of-sight formula by integrating the Boltzmann equation along a {\it perturbed} geodesic (curve) instead of a background geodesic (line).
In this approach, 
the separation of the gravitational and intrinsic effects are manifest. 
This approach can be considered as a generalization of the remapping approach of CMB lensing, 
where all the gravitational effects can be treated on the same footing.
}

\begin{flushright}
YITP-14-70
\end{flushright}

\maketitle

\section{Introduction}\label{sec:introduction}

The anisotropies of the cosmic microwave background (CMB) has been a rich source of information on the early stage of the universe. 
Their precise measurements have been playing a central role to confirm the standard $\Lambda$CDM cosmological model and to test the inflationary paradigm (e.g., \cite{Hinshaw:2012aka, Hou:2012xq, Sievers:2013ica, Calabrese:2013jyk, Ade:2013zuv, Ade:2013uln}).
In addition to its large amount of observable data, 
its utility as a probe of the early universe is attributed to calculability. 
The basic theory describing the CMB is well understood. 
Furthermore, 
because of the small amplitudes of the anisotropies, 
it can be analyzed with high accuracy by using the linear theory of perturbations \cite{Kodama:1985bj, Mukhanov:1990me}.

As measurements become more precise, 
tools are also required to be refined.
There are various small nonlinear effects that do not arise within the linear theory. 
They can contaminate signals from inflation 
and also give new tools to probe the late-time evolution of the universe.
One important example is the weak gravitational lensing of the CMB anisotropies (see e.g. Ref. \cite{Lewis:2006fu} for a review), 
which produces detectable levels of the bispectrum  \cite{Ade:2013uln} and the B-mode polarization \cite{Hanson:2013hsb, Ade:2013hjl, Ade:2013gez, Ade:2014afa} 
even when they are primordially absent.
Therefore, 
their contributions should be correctly taken into account to extract information on inflation from data.
In addition, 
its precise measurements can also help us to understand physics relevant to the late-time evolution of the universe such as dark energy and neutrino mass.
In any of these cases, 
it is important to correctly understand how the secondary nonlinear effects can affect the observed CMB anisotropies.

The leading nonlinear effects can be treated by expanding the evolution equations up to the second-order terms in perturbations.
The Boltzmann equation including the second-order terms was written down in Refs. \cite{Bartolo:2006cu, Bartolo:2006fj, Pitrou:2008hy, Pitrou:2008ut, Beneke:2010eg} and its gauge issue was discussed in Refs. \cite{Pitrou:2007jy, Naruko:2013aaa}. 
Their impact to the bispectrum was partially estimated in Refs. \cite{Khatri:2008kb, Bartolo:2008sg, Senatore:2008vi, Senatore:2008wk, Nitta:2009jp, Boubekeur:2009uk, Gao:2010ti, Creminelli:2011sq, Bartolo:2011wb}.
The full calculation was first accomplished in Ref. \cite{Pitrou:2010sn} and, subsequently, numerical codes to solve the full second-order Boltzmann equation were developed by several groups \cite{Su:2012gt, Huang:2012ub, Pettinari:2013he, Huang:2013qua}.

In the first-order case, 
the line-of-sight integration method \cite{Seljak:1996is} plays an important role to solve the Boltzmann equation with high accuracy and less computational time.
The direct integration of the Boltzmann equation is hard to be numerically performed.
For it, 
one should solve a large number of coupled differential equations 
because the multipole moments of the temperature anisotropies evolve interdependently in the free-streaming regime. 
On the other hand, 
the line-of-sight integration method enables us to efficiently calculate the observed temperature anisotropies 
by evaluating independent integration along a background geodesic for different multipoles.
In this approach, 
the observed temperature anisotropies are written as an integral over the product of a source term and a geometrical term, which represent the effects of collisions and propagation, respectively.
The source term depends only on a few low multipole moments of the temperature anisotropies in the last scattering and reionization epochs.
The information on the growth of the higher multipole moments in the free-streaming regime is encoded in the geometrical term. 
Because the geometrical term is written in terms of a known function,
it is not necessary to solve coupled differential equations to know its multipole moments.

However, 
it is not straightforward to extend the line-of-sight formula to second order. 
In contrast to the first-order case, 
the gravitational collision term, which is induced from the redshift, time-delay, and lensing effects, depend on the high multipole moments of the temperature anisotropies in the free-streaming regime. 
Hence, at second order, 
the source term also depends on the high multipole moments.
Since they appear with the metric perturbations, 
it is possible to evaluate them by using the first-order equations.
However, as first explained in Ref. \cite{Huang:2012ub},
it is still impractical to use the line-of-sight formula for the gravitational collision term because the multipole expansion of the product of the source term and geometrical term now introduces an infinite sum over multipoles.
Instead, the lensing effect, which gives the major contribution, has been included as remapping of the observed temperature anisotropies as done in its standard treatment \cite{Lewis:2006fu}.

A direct approach to the second-order gravitational effects from the Boltzmann equation was first developed by Huang \& Vernizzi \cite{Huang:2012ub, Huang:2013qua}.
They showed that the redshift effects can be separated from the source term by introducing a new variable for the temperature anisotropies and the other lensing and time-delay effects by performing integration by parts for the line-of-sight integral. 
The first method was also extended to the polarization in Ref. \cite{Fidler:2014oda} 
and the induced effects were estimated for the B mode \cite{Fidler:2014oda} and the bispectrum \cite{Pettinari:2014iha}.
The relation between the direct and remapping approaches to the lensing effect was also clarified by Su \& Lim \cite{Su:2014mga} by iteratively solving the Boltzmann equation.

 In this paper, we present another approach to the gravitational effects based on the Boltzmann equation, where the separation of the gravitational effects is easier to be seen.
The important fact for our approach is that the intensity of photons is conserved even when their trajectories are bent by the metric perturbations 
in the absence of a non-gravitational collision term (Liouville's theorem in curved spacetime \cite{Misner:1974qy}).
Based on this fact, 
we integrate the Boltzmann equation along a {\it perturbed} geodesic instead of a background one.
Then, we derive a mapping formula that relates the observed intensity to non-gravitational scattering sources.
In this formula, the gravitational effects appear as changes in the mapping between the coordinates of the observation and those of the sources.
This mapping formula can be considered as a generalization of the remapping formula in the standard treatment of CMB lensing.
Expanding this mapping formula up to second order, 
we demonstrate that one can directly obtain the second-order line-of-sight formula where the gravitational effects are separated from the non-gravitational source term, 
which depends only on a few low multipole moments of the intensity.
This approach provides a way to treat all the gravitational effects, i.e. redshift, time delay, and lensing, on the same footing.

The organization of this paper is as follows.
After providing the definitions for basic quantities in Sec. \ref{sec:def}, 
we review how the line-of-sight integration method simplifies the calculation at first order and why it is difficult to extend it to second order in more detail in Sec \ref{sec:pb}. 
These sections are used to introduce our notation.
We start the main discussion from Sec \ref{sec:LOS_df}.
In Sec. \ref{sec:LOS_df}, 
we derive the mapping formula and then the second-order line-of-sight formula for the observed intensity from it.
It is also shown in this section how the derived line-of-sight formula simplifies the computation of the bispectrum.
In Sec. \ref{sec:LOS_m}, 
we present the line-of-sight formula for the brightness and show that it reproduces the well-known result at first order.
We also discuss the relation between our approach and the remapping approach in Sec. \ref{sec:remapping}.
Finally, the summary of this paper is presented in Sec. \ref{sec:summary}.

\section{Definitions}\label{sec:def}
This section is devoted to provide the definitions for basic quantities used in this paper. 
Those readers primarily interested in the derivation of the line-of-sight formula can skip this section 
and return when necessary.

\subsection{Metric}\label{ss:g}
We write the metric in the ADM form \cite{Arnowitt:1962hi}:
	\begin{align}\label{def:metric}
		{\rm d}s^2 = a(\eta)^2\left[-e^{2\Psi}{\rm d}\eta^2 + \gamma_{ij}({\rm d}x^i + \omega^i{\rm d}\eta)({\rm d}x^j + \omega^j{\rm d}\eta) \right],
	\end{align}
where
	\begin{align}
		[\ln {\bs \gamma}]_{ij} \equiv 2h_{ij} \equiv 2\Phi\delta_{ij} + 2\chi_{ij}, \label{def:tensor}
	\end{align}
and impose the gauge conditions ${\omega^{i}}_{,i}=0$ and $\chi^i_i={\chi_i^j}_{,j}=0$. 
Here, 
the indices for the perturbations are lowered and raised with the Kronecker's delta. 
The logarithm of the spatial metric should be understood to be a function of a matrix,  
which is formally defined by its power series expansion. 
Hereafter, 
we use the bold symbol when an operation is understood as a matrix.
Then, 
the spatial metric is expanded in terms of $h_{ij}$ as,
	\begin{align}
		\gamma_{ij} &= \delta_{ij} + 2h_{ij} + 2h_i^k h_{kj} + \cdots  \\
		&= e^{2\Phi}(\delta_{ij} + 2\chi_{ij} + 2\chi_i^k \chi_{kj} + \cdots) . \label{eq:expand_gamma}
	\end{align} 
In this paper, 
we assume that the vector modes $\omega^i$ are of second order 
but do not for the tensor modes $\chi_{ij}$. 
Then, each metric perturbations are expanded as,
	\begin{align}
		&\Psi = \Psi^{\pI} + \Psi^{\pII} + \cdots, \quad \Phi = \Phi^{\pI} + \Phi^{\pII} + \cdots, \\
		&\omega_i \equiv  \delta_{ij}\omega^j =  \omega_i^{\pII} + \cdots, \quad \chi_{ij} = \chi_{ij}^{\pI} + \chi_{ij}^{\pII} + \cdots ,
	\end{align}
where the Greek numbers with brackets represent the order of the perturbative expansion. 
\footnote{We also use the bracketed indices for the tetrad components.
However, if a Greek number appears in brackets, it always represents the order of the perturbative expansion in this paper.}
The metric perturbations at each order are determined by solving the Einstein's equations at the corresponding order.
We also use a symbol with a bar to denote that it is estimated in the background.

Our definitions of the tensor modes are different from the standard one in the conformal Poisson gauge.
\footnote{This definition of the tensor mode is used in Ref. \cite{Maldacena:2002vr, Gao:2011vs, Gao:2012ib} when the non-Gaussianities are calculated, although the different time slicing is used there. }
By introducing the tensor modes in this way, 
extra quadratic terms do not appear in the resultant formulae. 
Since the estimation of such quadratic terms require the convolution of Fourier modes 
and increases computational time, 
the definition (\ref{def:tensor}) is employed in this paper.
For the difference between our definition and the standard one, 
see Appendix \ref{sec:metric_def}.

\subsection{Momentum of a photon}\label{ss:q}
We use the conformal momentum of a photon in the inertial frame $q^{(a)}$, which is defined as
	\begin{align}\label{def:qa}
		q^{(a)} \equiv a\tetrad{a}{\mu}p^{\mu},
	\end{align}
using the scale factor $a$, the tetrad $\tetrad{a}{\mu}$, and the momentum in the coordinate frame $p^{\mu}$. 
The tetrad and its inverse are respectively given by
	\begin{align}\label{eq:tetrad}
		\tetrad{0}{\mu} = a e^{\Psi}\delta^0_\mu, \quad \tetrad{i}{\mu}= a [e^{\boldsymbol h}]^i_j(\omega^j \delta^0_\mu + \delta^j_\mu),
	\end{align}
and 
	\begin{align}\label{eq:itetrad}
		\itetrad{0}{\mu} = \frac{e^{-\Psi}}{a}(\delta^\mu_0-\omega^i\delta^\mu_i), \quad \itetrad{i}{\mu} = \frac{1}{a}[e^{-{\boldsymbol h}}]^j_i \delta^\mu_j,
	\end{align}
modulo the local Lorentz transformations. 
They are chosen so that the time-like vector $\itetrad{0}{\mu}$ is orthogonal to the constant time hypersurfaces 
and the spatial vector $\itetrad{i}{\mu}$ are parallel to the coordinate basis vectors in the background. 
This frame was employed in Ref. \cite{Naruko:2013aaa}.
\footnote{On the other hand, the tetrad is chosen in Ref. \cite{Beneke:2010eg} so that $\itetrad{0}{\mu} \propto \delta^{\mu}_{0}$, which corresponds to a stationary observer in the coordinate system (\ref{def:metric}), $x^i = \text{const}$.}

In this paper, we mainly represent the momentum $q^{(a)}$ by its magnitude $q$ and direction $n^{(i)}$,
	\begin{align}\label{def:qni}
		q \equiv \sqrt{\delta_{(i)(j)}q^{(i)}q^{(j)}}, \quad n^{(i)} \equiv \frac{q^{(i)}}{q}.
	\end{align}
For brevity, we also denote the coordinates in the phase space $(x^i, q^{(a)})$ by $z^{(A)}$.

\subsection{Distribution function}\label{ss:f}

 The distribution of photons with polarization is represented by a tensor-valued distribution function $f_{\mu\nu}$ \cite{Pitrou:2008hy, Pitrou:2008ut, Beneke:2010eg, Naruko:2013aaa}. 
 The symmetric tensor $f_{\mu\nu}$ is defined on the hypersurface orthogonal to the observer's velocity $\itetrad{0}{\mu}$ and the direction of a photon $n^{\mu} \equiv \itetrad{i}{\mu}n^{(i)}$:
 	\begin{align}\label{eq:projection}
		{S_{\rho}}^{\mu}f_{\mu\nu} = f_{\rho\nu}, \quad {S_{\rho}}^{\nu}f_{\mu\nu} = f_{\mu\rho},
	\end{align}
 where $S_{\mu\nu}$ is the projection operator (screen projector),
 	\begin{align}\label{def:projection}
		S_{\mu\nu} \equiv g_{\mu\nu} + \tetrad{0}{\mu}\tetrad{0}{\nu} - n_{\mu}n_{\nu}.
	\end{align}
In a latter part, 
we also use the tetrad components of the projection operator,
 	\begin{align}\label{def:projection_ij}
		S_{(i)(j)} = \delta_{(i)(j)} - n_{(i)}n_{(j)}.
	\end{align}
When it is necessary to show its $n^{(i)}$-dependence explicitly, we will denote it as $S_{(i)(j)}(n^{(i)})$. 
Note that the distribution function is defined without gauge ambiguity but depend on the choice of the observer's frame.
See Ref. \cite{Naruko:2013aaa} for a discussion on how the observed temperature anisotropies are affected when the observer frame is changed.

 The four degrees of freedom of $f_{\mu\nu}$ can be extracted by decomposing it into a trace part $I$, a symmetric traceless part $P_{\mu\nu}$, and an antisymmetric part $V$ as,
 	\begin{align}\label{eq:decomposition}
		f_{\mu\nu} = \frac{1}{2}IS_{\mu\nu} + P_{\mu\nu} + \frac{i}{2}\epsilon_{\rho\mu\nu\sigma}\itetrad{0}{\rho}n^{\sigma}V,
	\end{align}
where $\epsilon_{\rho\mu\nu\sigma}$ is a completely antisymmetric tensor. 
Here, $I$, $P_{\mu\nu}$, and $V$ correspond to intensity, linear polarisation, and circular polarization, respectively.
\footnote{Note that the normalization for the intensity $I$ is different from that for $f_I$ in Ref. \cite{Beneke:2010eg} by a factor of two.}
In this paper, we only deal with the intensity $I$.

\section{The problem and its background}\label{sec:pb}

\subsection{Boltzmann/brightness equation}\label{ss:bbe}
To know the intensity $I$ of CMB photons coming from a direction $n_{\rm o}^{(i)} \equiv -n^{(i)}$ with comoving momentum $q$, 
one should solve the Boltzmann equation, 
	\begin{align}
		\dif{}{\eta}I(\eta, x^i, q, n^{(i)}) &\equiv \left(\pdif{}{\eta} + \dif{x^i}{\eta}\pdif{}{x^i} + \dif{q}{\eta}\pdif{}{q} + \dif{n^{(i)}}{\eta}\pdif{}{n^{(i)}}\right)I(\eta, x^i, q, n^{(i)}) \nm \\
		 &= {\mathfrak C}[I;\eta, x^i, q, n^{(i)}] , \label{eq:boltzmann_0}
	\end{align}
where, ${\mathfrak C}[I;\eta, x^i, q, n^{(i)}]$ represents the collision term for Compton scattering, 
which can be read from Ref. \cite{Beneke:2010eg} up to second order. 
\footnote{
Unlike Ref. \cite{Beneke:2010eg}, we included the factor $1/ap^0$ in the definition of the collision term, which appears when one rewrites the affine parameter in the Liouville term to the conformal time, $\eta$. 
}

Instead of treating the full $q$-dependence of the intensity, 
we usually consider the evolution of the brightness, the third moment of the intensity,
	\begin{align}\label{def:brightness}
		B(\eta, x^i,n^{(i)}) \equiv \int\!{\rm d}q~ q^3 I(\eta, x^i,q,n^{(i)}) ,
	\end{align}
because the $q$-dependence of the intensity can be well approximated by a Planck distribution, 
which can be characterized by a single parameter, i.e. temperature 
(See Appendix \ref{sec:bts} for the relation between the brightness and temperature).
Introducing its fraction to the background brightness by
	\begin{align}\label{def:delta}
		1 + \Delta \equiv \frac{B}{\bar{B}} ,
	\end{align}
the Boltzmann equation (\ref{eq:boltzmann_0}) reduces to a six-dimensional partial differential equation,
	\begin{align}\label{eq:brightness}
		\dot{\Delta} + n^{(i)}\pd_i \Delta = {\mathfrak C}^{\Delta}[\Delta] + {\mathfrak D}^{\Delta}[\Delta] ,
	\end{align}
where ${\mathfrak C}^{\Delta}$ represents the contribution from the collision term,
	\begin{align}\label{def:collision_delta}
		{\mathfrak C}^{\Delta}[\Delta] \equiv \frac{1}{\bar{B}}\int\!{\rm d}q~ q^3 {\mathfrak C}[I] ,
	\end{align}
and ${\mathfrak D}^{\Delta}$ from the Liouville term,
	\begin{align}\label{def:liouville_delta}
		 {\mathfrak D}^{\Delta}[\Delta] \equiv \left[ 4D^q - D^i \pd_i - D^{n^{(i)}}\pd_{n^{(i)}} \right](1+\Delta) .
	\end{align}
The ${\mathfrak D}^{\Delta}$ term can be considered as the collision term from gravitational scattering. 
Hence, we call it the gravitational collision term in this paper. 
The dot denotes the partial derivative with respect to the conformal time, $\eta$. 
In addition, $D^q$, $D^i$, and $D^{n^{(i)}}$ are terms in the photon geodesic equations dependent on the metric perturbations, whose explicit forms are given by
	\begin{align}
		D^q(\eta, x^i, n^{(i)}) \equiv -(1+\Psi-\Phi)\Psi_{,j}n^{(j)} + \chi^j_i \Psi_{,j} n^{(i)} - \dot{\Phi} - \left(\omega_{i,j} + \dot{\chi}_{ij}\right)n^{(i)}n^{(j)} ,
	\end{align}
and
	\begin{align}
		D^i(\eta, x^i, n^{(i)}) &\equiv (\Psi-\Phi)n^{(i)} - \chi^i_j n^{(j)},  \\
		D^{n^{(i)}}(\eta, x^i, n^{(i)}) &\equiv - S^{(i)(j)} \left[(\Psi-\Phi)_{,j} + \dot{\chi}_{jk}n^{(k)} + (\chi_{jl,k}-\chi_{kl,j})n^{(k)}n^{(l)} \right],
	\end{align}
up to second order. 
Here, 
$S_{(i)(j)}$ is the screen projector defined in Eq. (\ref{def:projection_ij}).
Note that it is sufficient to evaluate $D^i$ and $D^{n^{(i)}}$ up to first order 
because they are multiplied by the first-order quantities, $\pd_i \Delta$ or $\pd_{n^{(i)}}\Delta$.  
Following Ref. \cite{Pettinari:2013he}, 
we refer to each contribution from ${\rm d}\ln q/{\rm d}\eta$, ${\rm d} x^i/{\rm d}\eta$, and ${\rm d} n^{(i)}/{\rm d}\eta$ as redshift, time delay, and lensing, respectively.

On the other hand, 
the other $x^i$- and $n^{(i)}$-dependences are treated by decomposing into Fourier modes and spherical harmonics, respectively:
	\begin{align}
		\Delta_{lm}(\eta, {\bs k}) \equiv i^l \sqrt{\frac{2l+1}{4\pi}} \int {\rm d}^2 n Y^{\ast}_{lm}({\bs n}) \int \!{\rm d}^3 x e^{i{\bs k}\cdot{\bs x}}\Delta(\eta, {\bs x}, {\bs n}) .
	\end{align}
Then, 
the fractional brightness $\Delta$ with comoving wavenumber ${\bs k}$ and multipole $(l,m)$ evolves as
	\begin{align}\label{eq:brightness_lm}
		\dot{\Delta}_{lm} + k\sum_{l'm'} {\cal M}^{mm'}_{ll'}\Delta_{l'm'} = {\mathfrak C}^{\Delta}_{lm} + {\mathfrak D}^{\Delta}_{lm} ,
	\end{align}
where 
	\begin{align}
		 {\cal M}^{mm'}_{ll'} \equiv  \frac{\sqrt{(l+1+m)(l+1-m)}}{2l+3}\delta_{(l+1) l'}\delta_{mm'} - \frac{\sqrt{(l+m)(l-m)}}{2l-1}\delta_{(l-1) l'}\delta_{mm'} .
	\end{align}
Here, the pole of the polar coordinates has been chosen to be the direction of ${\bs k}$. 
Although each Fourier mode evolves independently at linear order, 
the multipole moments couple with each other due to the second term in Eq. (\ref{eq:brightness_lm}), which represents the effect of the propagation. 
Therefore, 
in order to know the values of the multipole moments relevant to current CMB experiments, 
one should solve more than thousands of coupled differential equations simultaneously.
In the following subsections, 
we review how the line-of-sight integration method enables us to calculate the brightness at the present time without directly solving the brightness equation (\ref{eq:brightness_lm}) 
and why it cannot be straightforwardly applied to second order.

\subsection{Line-of-sight integration method: the first-order case}
At first order, the collision term is given by
	\begin{align}
		{\mathfrak C}^{\Delta} \equiv \dot{\tau}C^{\Delta},
	\end{align}
where 
	\begin{align}\label{eq:1stCTb}
		C^{\Delta}[\Delta; {\bs n}] \equiv \Delta({\bs n}) - \frac{3}{2}\int\frac{{\rm d}\Omega'}{4\pi}S^{(i)(j)}({\bs n})\Delta_{(i)(j)}({\bs n'}) - 4{\bs n}\cdot{\bs v}_e ,
	\end{align}
with $\Delta_{(i)(j)}$ being the fractional brightness defined for the tetrad component of the distribution function for polarized photons 
\footnote{Unlike Refs. \cite{Fidler:2014oda, Pettinari:2014iha}, 
we have inserted a factor of two in the definition 
because the intensity $I$ is defined as the trace of $f_{(i)(j)}$ in this paper. 
With $S^{(i)(j)}\Delta_{(i)(j)} = 2\Delta$, we get Eq. (\ref{def:delta}) from the trace of Eq. (\ref{def:delta_ij}).}
(see the subsection \ref{ss:f}):
	\begin{align}\label{def:delta_ij}
		S_{(i)(j)}({\bs n}') + \Delta_{(i)(j)}({\bs n}') \equiv \frac{2}{\bar{B}}\left(\int\!{\rm d}q~ q^3 f_{(i)(j)}(q,{\bs n}')\right) .
	\end{align}
Here, we have suppressed the argument of the spacetime coordinates for brevity.
The vector ${\bs v}_e$ is the electron bulk velocity and $\tau$ the optical depth of photons defined through
	\begin{align}
		\dot{\tau} \equiv  \left(\pdif{}{\eta} + \dif{x^i}{\eta}\pdif{}{x^i} \right)\tau= -n_e\sigma_T a ,
	\end{align}
in terms of the electron number density $n_e$ and the Thomson scattering cross section $\sigma_T$, and $\tau=0$ at the present time. 
Here, contrary to its usage in the other part of this paper, 
the dot on $\tau$ denotes the derivative along a geodesic. 
The distinction is not important at first order, 
where it is sufficient to evaluate it in the background. 
However, 
it should be distinguished from the partial derivative at higher orders 
since its fluctuations can no longer be neglected.

Then, 
the brightness equation (\ref{eq:brightness}) can be rewritten as
	\begin{align}
		\left[\pdif{}{\eta} + n^{(i)}\pdif{}{x^i} \right]\left(e^{-\tau} \Delta \right) = e^{-\tau}{\mathfrak D}^{\Delta} + g_v S^{\Delta},
	\end{align}
where $g_v$ is the so-called visibility function,
	\begin{align}\label{def:gv}
		g_v \equiv -\dot{\tau} e^{-\tau} ,
	\end{align}
and we have introduced the source function $S^{\Delta} \equiv \Delta - C^{\Delta}$.
From Eq. (\ref{eq:1stCTb}),  its explicit form is
	\begin{align}\label{eq:sfb}
		S^{\Delta}[\Delta; {\bs n}] = \frac{3}{2}\int\frac{{\rm d}\Omega'}{4\pi}S^{(i)(j)}({\bs n})\Delta_{(i)(j)}({\bs n'}) + 4{\bs n}\cdot{\bs v}_e.
	\end{align}
The source function is written in terms of the multipole moments of the fractional brightness up to quadrupole. 
In addition, the series of its multipole expansion is finite, 
i.e. it terminates at quadrupole. 

This differential equation can be rewritten in a line-of-sight integral form \cite{Seljak:1996is},
	\begin{align}
		\Delta(\eta_0, x_0^i, n_0^{(i)}) = e^{-\tau(\eta_{\rm i})}\Delta(\eta_{\rm i}) 
		+ \int_{\eta_{\rm i}}^{\eta_0}{\rm d}\eta' \left( e^{-\tau}{\mathfrak D}^{\Delta} + g_v S^{\Delta} \right) ,
	\end{align}
where the functions in the RHS are evaluated along a photon geodesic in the background spacetime, 
	\begin{align}
		\bar{x}^i(\eta) &= n_0^{i}(\eta-\eta_0) + x_0^i , \\
		\bar{n}^{(i)}(\eta) &= n_0^{(i)} .
	\end{align}
If the initial time $\eta_{\rm i}$ is taken sufficiently before the recombination, 
the optical depth at this time is sufficiently large. 
Therefore, 
we can simplify the expression as,	
	\begin{align}\label{eq:LOSbg_x}
		\Delta(\eta_0, x_0^i,n_0^{(i)}) = \int_{0}^{\eta_0}{\rm d}\eta' \left( e^{-\tau}{\mathfrak D}^{\Delta} + g_v S^{\Delta} \right) ,
	\end{align}
or working in Fourier space, 
	\begin{align}\label{eq:LOSbg}
		\Delta(\eta_0, k^i,n_0^{(i)}) = \int_{0}^{\eta_0}{\rm d}\eta' \left( e^{-\tau}{\mathfrak D}^{\Delta} + g_v S^{\Delta} \right) e^{-i{\bs k}\cdot {\bs n}_0(\eta'-\eta_0)} ,
	\end{align}
where, in the RHS of Eq. (\ref{eq:LOSbg}), we have used the same symbols to denote the corresponding Fourier modes by abuse of notation.
At first order, 
the gravitational collision term ${\mathfrak D}^{\Delta}$ does not depend on $\Delta$,
	\begin{align}
		{\mathfrak D}^{\Delta\pI} = 4D^{q\pI} .
	\end{align}
Thus, 
 the fractional brightness $\Delta$ at the present time can be written as
	\begin{align}\label{eq:LOS1st}
		\Delta(\eta_0, k^i,n_0^{(i)}) = \int_{0}^{\eta_0}{\rm d}\eta' \left( 4e^{-\tau}D^{q\pI} + \bar{g}_v S^{\Delta(I)} \right)e^{-i{\bs k}\cdot {\bs n}_0(\eta'-\eta_0)} ,
	\end{align}
at first order. 
It is written as an integral over the product of a source term, the terms in the bracket, and a geometrical term, $e^{-i{\bs k}\cdot {\bs n}_0(\eta'-\eta_0)}$.
This equation is not solved for $\Delta$. 
However,
it determines the value of $\Delta$ at the present time from its low multipole moments 
because the source term only depends on the multipole moments up to quadrupole. 
In addition, no infinite sum over multipoles appears in the multipole expansion of the integrand 
because the multipole expansion of the source term is finite. 
The information on the growth of the higher multipole moments in the free-streaming regime is entirely encoded in a known function, $e^{-i{\bs k}\cdot {\bs n}_0(\eta'-\eta_0)}$. 
Then, 
its multipole dependence is written in terms of the spherical Bessel function, $j_l[k(\eta_0-\eta')]$. 
Thus, 
the line-of-sight formula (\ref{eq:LOS1st}) calculates the brightness at the present time without solving the hierarchical equations (\ref{eq:brightness_lm}) for the higher multipole moments.
The brightness at each multipole can be evaluated independently by performing the integration (\ref{eq:LOS1st}).

\subsection{A difficulty at second order}\label{ss:difficulty}

Now, we present why the line-of-sight integration method cannot be straightforwardly applied to second order.
At second order, 
we encounter a problem in treating the gravitational collision term ${\mathfrak D}^{\Delta}$ in Eq. (\ref{eq:LOSbg}). 
In contrast to the first-order case, 
${\mathfrak D}^{\Delta}[\Delta;{\bs n}]$ depends on $\Delta({\bs n})$ at second order (See Eq. (\ref{def:liouville_delta})). 
Therefore, the series of its multipole expansion is infinite and all multipole moments of $\Delta$ appear in it.
\footnote{The Compton collision term ${\mathfrak C}^{\Delta}$ also depends on high multipole moments of $\Delta$ at second order \cite{Beneke:2010eg}. However, the dependence vanish in the frame comoving with the electrons.}
In addition, it is multiplied by $e^{-\tau}$ instead of the visibility function $g_v \equiv -\dot{\tau}e^{-\tau}$. 
Then, 
it contributes to the integral in the entire free-streaming regime,  
where high multipole moments grow to be large.
Because they are multiplied by the metric perturbations, 
one can evaluate them by solving the first-order equations.
However, it is still necessary to perform an infinite sum over multipoles for all scales and times to evaluate the product of the source term and geometrical term because now both of them contain an infinite number of multipole moments.
Therefore, 
it is impractical to use the line-of-sight formula (\ref{eq:LOSbg}) at second order.

This difficulty was first discussed and overcome in Refs. \cite{Huang:2012ub, Huang:2013qua}. 
To eliminate the high multipole moments in the gravitational collision term, 
they made two manipulations. 
Leaving more detailed descriptions in Appendix \ref{sec:Huang}, 
in summary, they showed that 
(1) the redshift term becomes independent of $\Delta$ when the brightness equation is rewritten in terms of the new variable $\Delta_{\rm HV} \equiv \ln(1+\Delta)$ and 
(2) the high multipole moments in the remaining lensing and time-delay terms can be eliminated by performing integration by parts and using the brightness equation (\ref{eq:brightness}).

\begin{figure}[tbp]
		\centering
		\includegraphics[width=.7\linewidth]{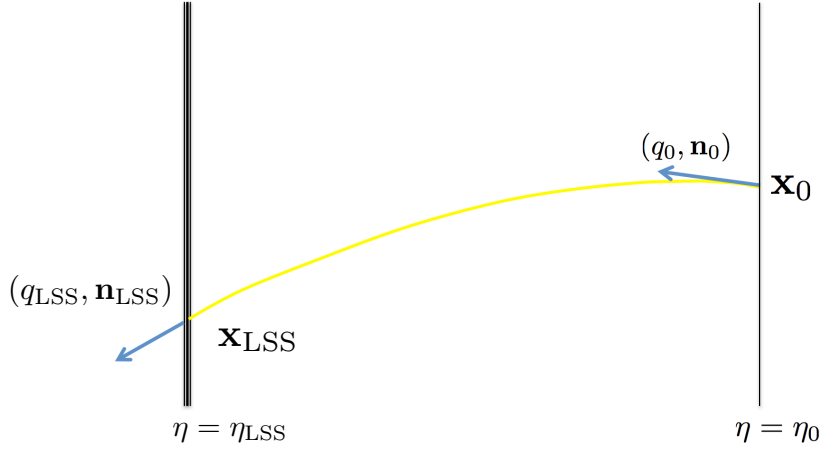}
		\caption{The intensity $I$ is conserved along a full geodesic in the free-streaming regime.}
		\label{fig:liouville}
	\end{figure}

In this approach, 
the gravitational interaction is treated as an external force that scatters the photons as in Eq. (\ref{eq:LOSbg}).
On the other hand, as in the original Boltzmann equation (\ref{eq:boltzmann_0}), 
its effects can be included geometrically in the Liouville term as changes of a geodesic.
It only changes the motion of the photons and does not the intensity.
This fact is known as the Liouville's theorem in curved spacetime \cite{Misner:1974qy}, 
which states conservation of the intensity along a {\it full} geodesic,
	\begin{align}\label{eq:conservation}
		I(\eta_0, x_0^i, q_0, n_0^{(i)}) = I(\eta_{\rm LSS}, x_{\rm LSS}^i, q_{\rm LSS}, n_{\rm LSS}^{(i)}) ,
	\end{align}
where two points $(\eta_0, x_0^i, q_0, n_0^{(i)})$ and $(\eta_{\rm LSS}, x_{\rm LSS}^i, q_{\rm LSS}, n_{\rm LSS}^{(i)}) $ are connected by a full geodesic as depicted in Fig. \ref{fig:liouville}. 
The intensity in the RHS gives a solution to the Boltzmann equation (\ref{eq:boltzmann_0}) 
when the non-gravitational collision term is absent, ${\mathfrak C}=0$ . 
Therefore, the Boltzmann equation can be solved once a solution to the geodesic equations is obtained.
It is not necessary to solve any evolution equations for multipole moments.

Starting from this consideration, 
in the next section, 
we introduce a new approach to derive a line-of-sight formula, 
where the Boltzmann equation is integrated along a {\it full} geodesic instead of a background one.
In this approach, one can automatically get expressions of a line-of-sight formula 
where the gravitational collision term does not appear in the integrand at all orders.
In the derivation, 
we will use the intensity instead of the brightness 
since it has a good property that it is conserved along a full geodesic.
This is not mandatory 
but will be helpful to clarify the discussion. 
A practical advantage to use the intensity is that one can also get line-of-sight formulae for spectral distortions 
since we have not integrated the $q$-dependence. 
In addition, 
it is expected that the extension to polarization will be easier in this approach
since the Boltzmann equation for polarization has a similar structure.  

\section{Curve-of-sight formulae}\label{sec:LOS_df}

\subsection{Mapping formula}\label{ss:mapping}

We start from the Boltzmann equation (\ref{eq:boltzmann_0}),
	\begin{align}
		\dif{}{\eta}I(\eta, z^{(A)}) &\equiv \left(\pdif{}{\eta} + \dif{z^{(A)}}{\eta}\pdif{}{z^{(A)}}\right)I(\eta, z^{(A)}) \nm  \\
		 &= \dot{\tau} C[I;\eta, z^{(A)}] , \label{eq:boltzmann_i}
	\end{align}
where, for brevity, we have denoted the coordinates in the phase space $(x^i, q, n^{(i)})$ by $z^{(A)}$.
As in the first-order case, 
we subtract $\dot{\tau}I$ in the both hand side. 
Then, 
the equation (\ref{eq:boltzmann_i}) can be rewritten as,
	\begin{align}
		\dif{}{\eta}\left[e^{-\tau}I(\eta, z^{(A)})\right] &= g_v(\eta')\left\{I(\eta, z^{(A)}) - C[I;\eta, z^{(A)}] \right\} \nm \\
		&\equiv g_v(\eta')S[I;\eta, z^{(A)}]  \equiv {\mathfrak S}[I;\eta, z^{(A)}] , \label{eq:boltzmann_ii}
	\end{align}
using the visibility function (\ref{def:gv}). 

Instead of integrating the Boltzmann equation (\ref{eq:boltzmann_ii}) along a background geodesic, 
we integrate it along a full geodesic. 
Then, 
the Boltzmann equation can be formally rewritten in an integral form as,
	\begin{align}\label{eq:LOSi}
		I(\eta_0, z_0^{(A)}) &= e^{-\tau(\eta_{\rm i})}I(\eta_{\rm i}, z^{(A)}(\eta_{\rm i};\eta_0, z_0^{(A)})) + \int_{\eta_{\rm i}}^{\eta_0}\!{\rm d}\eta' {\mathfrak S}[I;\eta', z^{(A)}(\eta';\eta_0, z_0^{(A)})],
	\end{align}
where $z^{(A)}(\eta;\eta_0, z_0^{(A)})$ is a solution of the full geodesic equations that satisfies a condition $z^{(A)} = z_0^{(A)}$ at the present time, $\eta=\eta_0$. 
 Note that the optical depth is also estimated along a full geodesic,
	\begin{align}
		\tau(\eta') = \int_{\eta'}^{\eta_0}\!{\rm d}\eta_1 n_e(\eta_1, x^i(\eta_1;\eta_0,z_0^{(A)}))\sigma_T a(\eta_1) .
	\end{align}
We can also define the optical depth through the electron number density in the background as,
	\begin{align}
		\bar{\tau}(\eta') = \int_{\eta'}^{\eta_0}\!{\rm d}\eta_1 \bar{n}_e(\eta_1)\sigma_T a(\eta_1) ,
	\end{align}
and then $g_v = \bar{g}_v$.
In this case, the optical depth does not depend on a geodesics. 
As we will explain later, 
the latter definition is convenient to evaluate the gravitational effects at second order. 
Therefore, we employ the latter definition in this paper.
In a similar fashion to the first-order case, 
we can simplify the expression as,
	\begin{align}
		I(\eta_0, z_0^{(A)}) &= \int_{0}^{\eta_0}\!{\rm d}\eta'  {\mathfrak S}[I;\eta', z^{(A)}(\eta';\eta_0, z_0^{(A)})] , \nm \\
		&= \int_{0}^{\eta_0}\!{\rm d}\eta'  \bar{g}_v(\eta')S[I;\eta', z^{(A)}(\eta';\eta_0, z_0^{(A)})] , \label{eq:LOSi2}
	\end{align}
by using $\tau(\eta_{\rm i}) \gg 1$ when the initial time $\eta_{\rm i}$ is taken sufficiently before the recombination. 
This integral form of the intensity along a full geodesic was obtained including polarization in Ref. \cite{Challinor:2000as}.
It has the required property. 
In the RHS of Eq. (\ref{eq:LOSi2}), 
only the non-gravitational collision term appears.
The effect of the gravitational interaction is entirely encoded in the mapping $z_0^{(A)} \to z^{(A)}(\eta';\eta_0, z_0^{(A)})$ for each time slice where the visibility function $g_v(\eta')$ is non-zero.
In contrast to the first-order line-of-sight formula (\ref{eq:LOS1st}), 
information on the metric perturbations is required to know this mapping.
However, 
since the distribution of photons is irrelevant to the evolution of the metric perturbations, 
\footnote{Note that we need to evaluate the multipole moments of the brightness up to $l \le 2$ even in the free-streaming regime 
because they affect the evolution of the metric perturbations through the energy-momentum tensor of the radiation (See Appendix \ref{sec:bts}). 
However, it will not be necessary to estimate them accurately 
because the radiation is subdominant, i.e.
its contributions are further suppressed by a factor of $\Omega_{\rm rad} < {\cal O}(0.1)$ in the free-streaming regime.}
the equation (\ref{eq:LOSi2}) gives a way to calculate the intensity at the present time without solving its evolution equation in the free-streaming regime.

Its relation to the conservation equation (\ref{eq:conservation}) becomes clearer when one neglects the non-gravitational scatterings after the last scattering and the width of the last scattering surface, i.e. $\bar{g}_v(\eta) \simeq \delta(\eta - \eta_{\rm LSS})$. 
In this case, 
the equation (\ref{eq:LOSi2}) becomes
	\begin{align}\label{eq:conservation_sharp}
		I(\eta_0, z_0^{(A)}) \simeq S(\eta_{\rm LSS}, z_{\rm LSS}^{(A)}) \equiv I(\eta_{\rm LSS}, z_{\rm LSS}^{(A)}) - C[I;\eta_{\rm LSS}, z_{\rm LSS}^{(A)}],
	\end{align}
where $(\eta_{\rm LSS}, z_{\rm LSS}^{(A)})$ represent a point where a full geodesic crosses the last scattering surface, 
which is determined by $z_{\rm LSS}^{(A)} = z^{(A)}(\eta_{\rm LSS};\eta_0, z_0^{(A)})$ as a function of $(\eta_0, z_0^{(A)})$. 
The equation (\ref{eq:conservation_sharp}) represents the conservation of the intensity (\ref{eq:conservation}) taking into account the effect of the collision term at the last scattering surface.

Now, 
we expand the RHS of Eq. (\ref{eq:LOSi2}) with respect to the perturbations. 
The sources of the perturbations can be categorized into three types: 
(i) a deviation of a geodesic due to the metric perturbations $\delta z^{(A)}(\eta';\eta_0, z_0^{(A)})$, 
(ii) the perturbations in the electron distribution $\delta n_e$, ${\bs v}_e$, 
and (iii) intrinsic perturbations in the intensity. 
\footnote{This separation is not gauge invariant because the metric and matter perturbations mix with each other by a gauge transformation.}
Replacing $z^{(A)}(\eta';\eta_0, z_0^{(A)})$ by its background counterpart $\bar{z}^{(A)}(\eta';\eta_0, z_0^{(A)})$ in Eq. (\ref{eq:LOSi2}), 
the intensity with the latter two non-gravitational effects is given by
	\begin{align}\label{def:ILSS}
		I_{\rm LSS}(\eta_0, z_0^{(A)}) \equiv \int_{0}^{\eta_0}\!{\rm d}\eta'  {\mathfrak S}[I;\eta', \bar{z}^{(A)}(\eta';\eta_0, z_0^{(A)})] , 
	\end{align}
where explicitly
	\begin{align}
		\bar{x}^i(\eta';\eta_0, x^i_0, q^{(a)}_0) &= n^{(i)}_0(\eta'-\eta_0) + x^i_0, \label{eq:SGE_BGx}\\
		\bar{q}(\eta';\eta_0, x^i_0, q^{(a)}_0) &= q_0, \label{eq:SGE_BGq} \\
		\bar{n}^{(i)}(\eta';\eta_0, x^i_0, q^{(a)}_0) &= n^{(i)}_0. \label{eq:SGE_BGn}
	\end{align}
The contributions $I_{\rm LSS}$ only contains perturbed quantities in the last scattering surface with corrections from scatterings in the reionization epoch.
Then, 
the contributions to the observed intensity from the gravitational effects, $\delta I_G$, can be represented by
\footnote{Strictly speaking, $\delta I_G$ is not contributions purely from the gravitational effects because the source function in the definition (\ref{eq:LOSig}) is the perturbed one.
However, since it vanishes in the absence of the metric perturbations, we call it the contributions from the gravitational effects.}
	\begin{align}
		\delta I_{\rm G}(\eta_0, z_0^{(A)}) &\equiv I(\eta_0, z_0^{(A)}) - I_{\rm LSS}(\eta_0, z_0^{(A)}) \\
		&= \int_{0}^{\eta_0}\!{\rm d}\eta' \left\{ {\mathfrak S}[I;\eta', z^{(A)}(\eta';\eta_0, z_0^{(A)})] - {\mathfrak S}[I;\eta', \bar{z}^{(A)}(\eta';\eta_0, z_0^{(A)})] \right\} . \label{eq:LOSig}
	\end{align}
Expanding it in terms of the deviation
	\begin{align}
		\delta z^{(A)}(\eta';\eta_0, z_0^{(A)}) \equiv z^{(A)}(\eta';\eta_0, z_0^{(A)}) - \bar{z}^{(A)}(\eta';\eta_0, z_0^{(A)}) ,
	\end{align}
we obtain a formula
	\begin{align}
		&\delta I_{\rm G}(\eta_0, z_0^{(A)}) = \nm \\ 
		&\quad \int_{0}^{\eta_0}\!{\rm d}\eta' \left\{ \left[\pdif{{\mathfrak S}}{z^{(A)}}\right]_{z^{(A)}=\bar{z}^{(A)}}\delta z^{(A)}  + \frac{1}{2}\left[\frac{\pd^2 {\mathfrak S}}{\pd z^{(A)}\pd z^{(B)}}\right]_{z^{(A)}=\bar{z}^{(A)}}\delta z^{(A)}\delta z^{(B)} + \cdots \right\}, \label{eq:mapping}
	\end{align}
which relates the intensity at the present time to the non-gravitational scattering sources. 
The terms in the integrand are written as the products of a source term, the derivatives of the source term, ${\mathfrak S}$, and the deviations of a geodesic, $\delta z^{(A)}$, evaluated at a position of a source. 
The perturbations in the intensity are given by its integral over the positions of all sources.
The source term is written only in terms of the non-gravitational collision term.
On the other hand, 
the deviation $\delta z^{(A)}$ only depends on the metric perturbations between the observer and sources.
Thus,  
the gravitational effects are separated from the intrinsic effects in Eq. (\ref{eq:mapping}). 
This is automatically guaranteed at all orders. 

The equation (\ref{eq:mapping}) can be considered as a generalization of the standard treatment of CMB lensing as remapping, 
where all the gravitational effects are treated on the same footing. 
We discuss the relation to the standard treatment of CMB lensing in more detail in Sec. \ref{sec:remapping}.

\subsection{Deviation of a geodesic}\label{ss:deviations}

 To evaluate the intensity at the present time from the mapping formula (\ref{eq:mapping}), 
 one should know how the metric perturbations change the mapping $z_0^{(A)} \to z^{(A)}(\eta';\eta_0, z_0^{(A)})$. 
Here, 
we give explicit forms of the deviation $\delta z^{(A)}$ in terms of the metric perturbations by solving the geodesic equations iteratively for them.

In the required accuracy, 
the geodesic equations are given by
	\begin{align}
		\dif{x^i}{\eta} &= n^{(i)} + D^i(\eta, x^i, n^{(i)}), \label{eq:geodesic2nd_x}\\
		\frac{1}{q}\dif{q}{\eta} &= D^q(\eta, x^i, n^{(i)}), \label{eq:geodesic2nd_q}\\
		\dif{n^{(i)}}{\eta} &= D^{n^{(i)}}(\eta, x^i, n^{(i)}) , \label{eq:geodesic2nd_n}
	\end{align}
where the functions $D^i$, $D^q$, and $D^{n^{(i)}}$ are same as those defined in the subsection \ref{ss:bbe}:
	\begin{align}
		D^i(\eta, x^i, n^{(i)}) &\equiv (\Psi-\Phi)n^{(i)} - \chi^i_j n^{(j)} ,\\
		D^q(\eta, x^i, n^{(i)}) &\equiv -(1+\Psi-\Phi)\Psi_{,j}n^{(j)} + \chi^j_i \Psi_{,j} n^{(i)} - \dot{\Phi} - \left(\omega_{i,j} + \dot{\chi}_{ij}\right)n^{(i)}n^{(j)} ,\\
		D^{n^{(i)}}(\eta, x^i, n^{(i)}) &\equiv - S^{(i)(j)}\left[(\Psi-\Phi)_{,j} + \dot{\chi}_{jk}n^{(k)} + (\chi_{jl,k}-\chi_{kl,j})n^{(k)}n^{(l)} \right]  .
	\end{align}
Here, 
we made evaluations up to second order only for $q$ 
because the coefficients of the series expansion for $\delta x_0^i$ and $\delta n_0^{(i)}$ in Eq. (\ref{eq:mapping}) have no zeroth-order terms. 
Note that the dots on the metric perturbations represent the partial derivatives with respect to $\eta$ 
unlike the definition of the dot on $\tau$.

\begin{figure}[tbp]
		\centering
		\includegraphics[width=.8\linewidth]{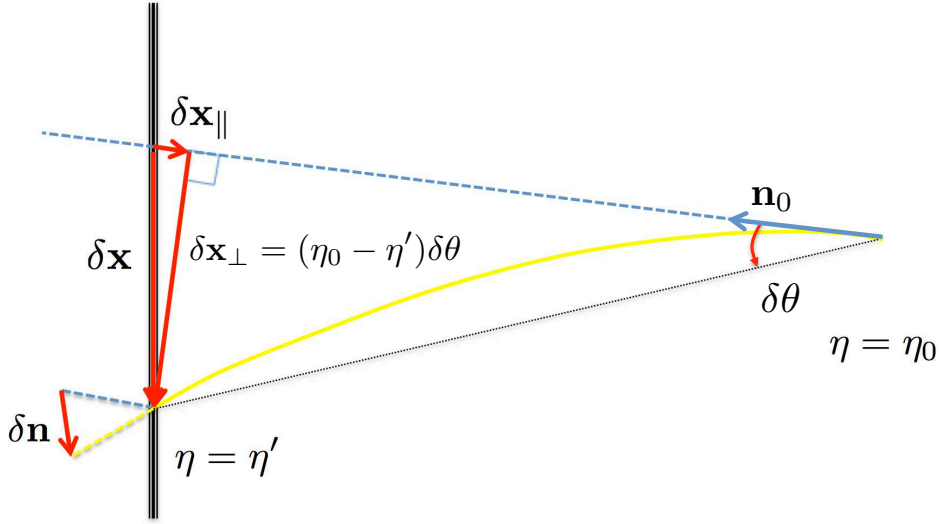}
		\caption{Illustration for the geometrical meanings of the deviations $\delta x^i$ and $\delta n^{(i)}$. The straight dashed line and curved solid line represent a background geodesic and a full geodesic, respectively. The components of the displacement $\delta x^i$ parallel and perpendicular to the line of sight correspond to time delay and lensing, respectively. On the other hand, $\delta n^{(i)}$ corresponds to deflection at a position of a source.}
		\label{fig:geometrical_effects}
	\end{figure}

 Since the geodesic equations for $x^i$  and $n^{(i)}$ do not depend on $q$, 
 we can solve them separately.
Up to first order, 
the deviations $\delta x^i$ and $\delta n^{(i)}$ should satisfy
	\begin{align}
		\dif{}{\eta}\left[ \delta x^i - \eta\delta n^{(i)} \right] &= D^i(\eta, \bar{x}^i, \bar{n}^{(i)}) - \eta D^{n^{(i)}}(\eta, \bar{x}^i, \bar{n}^{(i)}), \label{eq:geodesic2ndx} \\
		\dif{\delta n^{(i)}}{\eta} &= D^{n^{(i)}}(\eta, \bar{x}^i, \bar{n}^{(i)}), \label{eq:geodesic2ndn}
	\end{align}
with a condition that $(\delta x^i, \delta n^{(i)})$ vanishes at $\eta=\eta_0$. 
Using iteration, we can solve them as,
	\begin{align}
		\delta x^i(\eta') &= -\int_{\eta'}^{\eta_0}\!{\rm d}\eta_1 \left[ D^i(\eta_1, \bar{x}^i, \bar{n}^{(i)}) - (\eta_1 -\eta')D^{n^{(i)}}(\eta_1, \bar{x}^i, \bar{n}^{(i)}) \right], \label{eq:SGE_Px} \\
		\delta n^{(i)}(\eta') &= - \int_{\eta'}^{\eta_0}\!{\rm d}\eta_1 D^{n^{(i)}}(\eta_1, \bar{x}^i, \bar{n}^{(i)}) , \label{eq:SGE_Pn}
	\end{align}
in terms of the metric perturbations.
The displacement $\delta x^i$ can be decomposed as,
	\begin{align}
		\delta x^i = \delta x_{\parallel}n_{\rm o}^{(i)} + S_{\rm o}^{(i)(j)}\delta x_{\perp {(j)}} ,
	\end{align}
for the direction of the observation $n_{\rm o}^{(i)} \equiv -n_0^{(i)}$, where
	\begin{align}
		\delta x_{\parallel}(\eta') &\equiv -\int_{\eta'}^{\eta_0}\!{\rm d}\eta_1 n_{{\rm o}(i)}D^i(\eta_1, \bar{x}^i, \bar{n}^{(i)}) , \label{eq:parallel}\\
		\delta x_{\perp}^{(i)}(\eta') &\equiv \int_{\eta'}^{\eta_0}\!{\rm d}\eta_1 (\eta_1 -\eta')D^{n^{(i)}}(\eta_1, \bar{x}^i, \bar{n}^{(i)}) , \label{eq:orthogonal}
	\end{align}
and $S_{\rm o}^{(i)(j)} \equiv S^{(i)(j)}(n_{\rm o}^{(i)})$.
We illustrated the geometrical meaning of each term in Fig. \ref{fig:geometrical_effects}. 
The displacements $\delta x_{\parallel}$ and $\delta x_{\perp}^{(i)}$ correspond to time delay and lensing, respectively. 
As for $\delta n^{(i)}$, 
it corresponds to deflection at a position of a source. 
In fact, 
since the change in the observed angle $\delta \theta^{(i)}$ relates to $\delta x_{\perp}^{(i)}$ as $\delta x_{\perp}^{(i)} = (\eta_0-\eta')\delta \theta^{(i)}$, 
it can be expressed as
	\begin{align}\label{eq:angle}
		\delta \theta^{(i)} = \int_{\eta'}^{\eta_0}\!{\rm d}\eta_1 \frac{\eta_1 -\eta'}{\eta_0-\eta'}D^{n^{(i)}}(\eta_1, \bar{x}^i, \bar{n}^{(i)}).
	\end{align}
Then, 
replacing the spatial derivative $S^{(i)(j)}\pd_j$ in $D^{n^{(i)}}(\eta_1, \bar{x}^i, \bar{n}^{(i)})$ by the angular derivative $\pd_{n_{\rm o}^{(i)}}/(\eta_0-\eta_1)$,
the contribution from the scalar modes can be written as
	\begin{align}
		\delta \theta^{(i)} = \pd_{n_{\rm o}^{(i)}}\psi ,
	\end{align}
by using the lensing potential \cite{Lewis:2006fu},
	\begin{align}
		\psi \equiv -\int_{\eta'}^{\eta_0}\!{\rm d}\eta_1 \frac{\eta_1 -\eta'}{(\eta_0-\eta')(\eta_0-\eta_1)}(\Psi-\Phi) .
	\end{align}
The factor $(\eta_1-\eta')$ in Eq. (\ref{eq:orthogonal}) make the lensing contributions larger than those from the others. 
This factor arises because the geodesic equation (\ref{eq:geodesic2nd_x}) has the term that is not suppressed by the metric perturbations, $n^{(i)}$.
Then, deflection $\delta n^{(i)}$ cumulatively causes the displacement $\delta x^i$.

Provided the solutions for $\delta x^i$ and $\delta n^{(i)}$, 
we can find the solution for redshift, $\delta \ln q$.
Before writing down the solution, 
it is convenient to rewrite the redshift equation (\ref{eq:geodesic2nd_q}). 
By using Eq. (\ref{eq:geodesic2nd_x}), the first two terms in $D^q$ can be rewritten as
	\begin{align}
		-(1+\Psi-\Phi)\Psi_{,j}n^{(j)} + \chi^j_i \Psi_{,j} n^{(i)} &= -\Psi_{,j}\dif{x^i}{\eta} \nm \\
		&= - \dif{\Psi}{\eta} + \dot{\Psi} . \nm 
	\end{align}
Thus, 
we can get a more familiar expression for the redshift equation,
	\begin{align}\label{eq:SWISW}
		\dif{}{\eta}\left[\ln q + D^{\rm SW}(\eta, x^i) \right] &= D^{\rm ISW}(\eta, x^i, n^{(i)}) ,
	\end{align}
where
	\begin{align}
		D^{\rm SW}(\eta, x^i) &\equiv \Psi, \label{def:SW}\\
		D^{\rm ISW}(\eta, x^i, n^{(i)}) &\equiv (\Psi-\Phi)^{\cdot} - \left(\omega_{i,j} + \dot{\chi}_{ij}\right)n^{(i)}n^{(j)} . \label{def:ISW}
	\end{align}
The term $D^{\rm SW}$ corresponds to the Sachs-Wolfe (SW) effect and $D^{\rm ISW}$ the integrated Sachs-Wolfe (ISW) effect. 
As was mentioned in Ref. \cite{Pettinari:2013he}, 
this SW-ISW decomposition is useful in practice
because the time derivative of the gravitational potentials are much smaller than their spatial derivatives. 
In Appendix \ref{sec:GE}, 
we show how the SW-ISW decomposition can be generalized to higher orders.

Integrating (\ref{eq:SWISW}), 
we obtain
	\begin{align}
		\delta \ln q (\eta') = [\delta \ln q (\eta')]_{\rm SW} + [\delta \ln q (\eta')]_{\rm ISW} , \label{eq:SGE_Pq}
	\end{align}
where
	\begin{align}
		[\delta \ln q (\eta')]_{\rm SW} = - D^{\rm SW}(\eta', \bar{x}^i) - \delta x^i(\eta') \pd_i D^{\rm SW}(\eta', \bar{x}^i),\label{eq:SGE_PSW}
	\end{align}
and 
	\begin{align}
		[\delta \ln q (\eta')]_{\rm ISW} &= 
		- \int_{\eta'}^{\eta_0}\!{\rm d}\eta_1 \left[ D^{\rm ISW}(\eta_1, \bar{x}^i, \bar{n}^{(i)}) + \delta x^i(\eta_1)\pd_i D^{\rm ISW} + \delta n^{(i)}(\eta_1)\pd_{n^{(i)}}D^{\rm ISW} \right] , \label{eq:SGE_PISW}
	\end{align}
up to second order. 
The terms depending on $\delta x^i$ and $\delta n^{(i)}$ arise 
because redshift should be evaluated along a perturbed trajectory instead of a background one.
They are induced from the couplings between redshift and the other gravitational effects, i.e. time delay, lensing, and deflection.

\subsection{Line-of-sight formulae at second order}\label{ss:LOS2nd}

Now that we have known how the mapping $z_0^{(A)} \to z^{(A)}(\eta';\eta_0, z_0^{(A)})$ changes due to the metric perturbations, in this section, 
we write down explicit forms of line-of-sight formulae at second order from the mapping formula (\ref{eq:mapping}).

Note that, since the source function ${\mathfrak S}$ is multiplied by the deviation $\delta z^{(A)}$ in Eq. (\ref{eq:mapping}), 
it is sufficient to estimate it up to first order,
	\begin{align}\label{eq:1stCTs}
		{\mathfrak S}[I; {\bs n}] = \bar{g}_v \left[ \frac{3}{2}\int\frac{{\rm d}\Omega'}{4\pi}S^{(i)(j)}({\bs n})f_{(i)(j)}({\bs n'}) - ({\bs n}\cdot{\bs v}_e)\pd_{\ln q} \bar{I} \right] ,
	\end{align}
where $f_{(i)(j)}$ is the tetrad component of the distribution function for polarized photons defined in the subsection \ref{ss:f}.
Here, we have suppressed the arguments of ${\bs x}$ and $q$. 
The source function is only written in terms of the multipole moments of the distribution function up to quadrupole and the series of its multipole expansion terminates at quadrupole. 
As stated in the subsection \ref{ss:mapping}, here, 
we have defined the optical depth through the electron number density in the background. 
In general, this introduces the extra term in the source function, $e^{-\bar{\tau}}\delta \dot{\tau} C$, 
which depends on the higher-order multipole moments of the distribution function. 
However, it is sufficient to evaluate the function $C$ in the background, $\bar{C}=0$, for the first-order source function. 
Therefore, in the evaluation of the gravitational effects at second order, the extra term vanishes and then the higher-order multipole moments do not appear in the source function.

At second order, only four terms are relevant in the mapping formula (\ref{eq:mapping}):
\begin{align}
		\delta I_{\rm G}(\eta_0, z_0^{(A)}) \simeq \int_{0}^{\eta_0}\!{\rm d}\eta' \left\{ \left[\pdif{{\mathfrak S}}{x^i}\right]\delta x^i + \left[\pdif{{\mathfrak S}}{n^{(i)}}\right]\delta n^{(i)} + \left[\pdif{{\mathfrak S}}{\ln q}\right]\delta \ln q  + \frac{1}{2}\left[\frac{\pd^2 {\mathfrak S}}{\pd \ln q^2}\right](\delta \ln q)^2 \right\}. \label{eq:mapping2nd}
	\end{align}
From the discussion in the previous subsection, 
their geometrical meanings are clear.
The first term represents the lensing and time-delay effects and the second term the contribution from deflection at a position of a source. 
The third term includes the redshift effects like the Sachs-Wolfe (SW) and the integrated Sachs-Wolfe (ISW) effects as well as their coupling terms with the lensing, time-delay, and deflection effects.
Finally, the forth term represents the coupling terms between two redshift effects. 
Here, we evaluate them in order. \\

- {\it \large Lensing, time delay, and deflection}\\

 First, we consider the first and second terms in Eq. (\ref{eq:mapping2nd}). 
 As for the lensing, time-delay, and deflection terms, 
 the source terms are at least of first order. 
 Therefore, it is sufficient to perform the expansion up to first order for $\delta x^i$ and $\delta n^{(i)}$ and evaluate them in the first-order accuracy,
 	\begin{align}
		\delta I_{\rm G}^{(2)} \supset \int_{0}^{\eta_0}\!{\rm d}\eta' \left\{ \left[\pdif{{\mathfrak S}}{x^i}\right]^{\pI}[\delta x^i]^{\pI} + \left[\pdif{{\mathfrak S}}{n^{(i)}}\right]^{\pI}[\delta n^{(i)}]^{\pI} \right\} .
	\end{align}
Because of the statistical homogeneity of the fluctuations,
we can set the point of the observation as the origin, ${\bs x}_0 = {\bs o}$, without loss of generality.
Then, 
in terms of the Fourier modes, 
 these contributions have a similar form as,
 	\begin{align}
		\delta I_{\rm N} = \frac{1}{(2\pi)^6}\int{\rm d}^3k_1 \int_{0}^{\eta_0}\!{\rm d}\eta' {\mathfrak S}_{N}^{\pI}({\bs k}_1, \eta') \left[ \int{\rm d}^3k_2 \int_{\eta'}^{\eta_0}\!{\rm d}\eta_1 T^{N\pI}({\bs k}_2, \eta_1;\eta')e^{-i[{\bs k}_1(\eta_0-\eta')+{\bs k}_2(\eta_0-\eta_1)] \cdot {\bs n}_{\rm o}}  \right] , \nm \\ 
		(N={\rm L}, {\rm TD}, {\rm D}), \label{eq:IN}
	\end{align}
where the source function ${\mathfrak S}_N^{\pI}$ represents $k_i{\mathfrak S}^{\pI}$ for lensing (L) and time delay (TD), and $\pd_{n^{(i)}}{\mathfrak S}^{\pI}$ for deflection (D). 
Here, the index $i$ of ${\mathfrak S}_N^{\pI}$ should be contracted with the corresponding index of $T^{N\pI}$. 
The terms $T^{N\pI}~(N={\rm L}, {\rm TD}, {\rm D})$ are given by 
	\begin{align}
		T^{{\rm L}\pI}({\bs k}, \eta_1;\eta') &\equiv (\eta_1-\eta')\tilde{T}^{{\rm L}\pI}({\bs k}, \eta_1) \nm \\
		&\equiv (\eta_1-\eta')S_{\rm o}^{(i)(j)}\left[k_{j}(\Psi^{\pI}-\Phi^{\pI}) + i\dot{\chi}_{jk}^{\pI}n_{\rm o}^{(k)} + (k_{k}\chi_{jl}^{\pI}-k_{j}\chi_{kl}^{\pI})n_{\rm o}^{(k)}n_{\rm o}^{(l)} \right] , \\
		T^{{\rm TD}\pI}({\bs k}, \eta_1) &\equiv in_{{\rm o}}^{(i)}\left[(\Psi^{\pI}-\Phi^{\pI}) - \chi_{jk}^{\pI} n_{\rm o}^{(j)}n_{\rm o}^{(k)}\right] ,
	\end{align}
for the lensing and time-delay effects, and 
	\begin{align}
		T^{{\rm D}\pI}({\bs k}, \eta_1) &\equiv iS_{\rm o}^{(i)(j)}\left[k_{j}(\Psi^{\pI}-\Phi^{\pI}) + i\dot{\chi}_{jk}^{\pI}n_{\rm o}^{(k)} + (k_{k}\chi_{jl}^{\pI}-k_{j}\chi_{kl}^{\pI})n_{\rm o}^{(k)}n_{\rm o}^{(l)} \right] , 
	\end{align}
for the deflection effect. 
The dependence on $\eta'$ appears only for the lensing contributions. 
Note also that $\pd_{n^{(i)}}{\mathfrak S}$ contains a background term but it vanishes when contracted with $T^{{\rm D}}$. 
Therefore, 
it is sufficient to evaluate the perturbed quantities in ${\mathfrak S}_N$ and $T^N$ up to first order. 
These formulae correspond to Eq. (65) in Ref. \cite{Su:2014mga}.

These contributions to the intensity (\ref{eq:IN}) are written as an integral over the product of a source term, ${\mathfrak S}_N$, and a geometrical term, the term in the square brackets. 
The effect of the free streaming is entirely encoded in the geometrical term. 
In contrast to the first-order case (\ref{eq:LOS1st}), 
the geometrical term is no longer a known function.
However, it depends on ${\bs n}_0$ only through known functions, i.e. the products of ${\bs n}_{\rm o}$ and $e^{-i[{\bs k}_1(\eta_0-\eta')+{\bs k}_2(\eta_0-\eta_1)] \cdot {\bs n}_{\rm o}}$. 
Then,
its multipole dependence is written in terms of the spherical Bessel function. 
In the subsection \ref{ss:bispectrum}, 
as an example, we will show how it simplifies the computation of the bispectrum.

Since the source term is multiplied by the visibility function, ${\mathfrak S} \equiv g_v S$, 
the integration domain for $\eta'$ is limited to the last scattering and reionization epochs. 
The formula is linear in the source term. 
Therefore, the contributions from different epochs can be separately estimated.
For each, 
the geometrical term can be estimated by integrating the first-order metric perturbations with the spherical Bessel function between the observer and sources. 

The series expansion (\ref{eq:mapping}) for lensing $\delta x_{\perp}^{(i)}$ can be inaccurate for small scales, $k_1k_2(\eta_0 - \eta_{\rm LSS})^2 {\cal O}(|\delta g_{\mu\nu}|) > 1$. 
Here, ${\cal O}(|\delta g_{\mu\nu}|)$ represents terms of the order of the metric perturbations.
Although we presented line-of-sight formulae based on the series expansion (\ref{eq:mapping}) in this paper, 
it will be possible to make the estimation more accurate by developing the nonperturbative approach similar to that used in the standard treatment of CMB lensing \cite{Lewis:2006fu}. \\

- {\it \large Linear redshift} \\

 Next, we consider the third term in Eq. (\ref{eq:mapping2nd}).
 In the case of the redshift effects, 
 the source terms contain the background terms.
 Therefore, we need to perform the expansion up to second order for $\delta \ln q$ in Eq. (\ref{eq:mapping}).
 Among them, here, we evaluate the linear terms.
 
 Since the source term contains the background terms, 
 one needs to evaluate $\delta \ln q$ up to second order,
 	\begin{align}
		\delta I_{\rm G}^{\pII} \supset \int_{0}^{\eta_0}\!{\rm d}\eta' \left\{ \left[\pdif{{\mathfrak S}}{\ln q}\right]^{\pI}[\delta \ln q]^{\pI} + \left[\pdif{\mathfrak S}{\ln q}\right]^{\pO}[\delta \ln q]^{\pII} \right\} ,
	\end{align}
where the quantities with superscript $\pO$ indicates that they are evaluated in the background.
The first-order contributions can be evaluated as,
	\begin{align}
		\delta I_{{\rm SW}\pI} = \frac{1}{(2\pi)^6}\int{\rm d}^3k_1 \int_{0}^{\eta_0}\!{\rm d}\eta' \left[ \pd_{\ln q}{\mathfrak S}({\bs k}_1, \eta')\right]^{\pI} \left[ \int{\rm d}^3k_2 T^{{\rm SW}\pI}({\bs k}_2, \eta')e^{-i({\bs k}_1+{\bs k}_2) \cdot {\bs n}_{\rm o}(\eta_0-\eta')}  \right] ,\label{eq:ISW_I}
	\end{align}
for the SW effect, and
	\begin{align}
		\delta I_{{\rm ISW}\pI} = \frac{1}{(2\pi)^6}\int{\rm d}^3k_1 \int_{0}^{\eta_0}\!{\rm d}\eta' \left[ \pd_{\ln q}{\mathfrak S}({\bs k}_1, \eta') \right]^{\pI} \left[ \int{\rm d}^3k_2 \int_{\eta'}^{\eta_0}\!{\rm d}\eta_1 T^{{\rm ISW}\pI}({\bs k}_2, \eta_1)e^{-i[{\bs k}_1(\eta_0-\eta')+{\bs k}_2(\eta_0-\eta_1)] \cdot {\bs n}_{\rm o}}  \right] , \label{eq:IISW_I}
	\end{align}
for the ISW effect. 
Here, the terms $T^{{\rm SW}\pI}$ and $T^{{\rm ISW}\pI}$ are defined as,
 	\begin{align}
		T^{{\rm SW}\pI}({\bs k}, \eta') &\equiv -\Psi^{\pI}, \\
		T^{{\rm ISW}\pI}({\bs k}, \eta_1) &\equiv -\left[\Psi^{\pI}-\Phi^{\pI}\right]^{\cdot} + \dot{\chi}_{ij}^{\pI}n_{\rm o}^{(i)}n_{\rm o}^{(j)}. \label{eq:TISW_I}
	\end{align}
The contribution $\delta I_{{\rm ISW}\pI}$ can be evaluated in a way similar to the contributions (\ref{eq:IN}).

On the other hand, the second-order contributions are further separated into two terms:
	\begin{align}\label{eq:qIIsplit}
		[\delta \ln q]^{\pII} = [\delta \ln q]^{\pIIL} + [\delta \ln q]^{\pIIQ} ,	
	\end{align}
where the first and second terms are linear and quadratic in the metric perturbations respectively,
	\begin{align}
		[\delta \ln q]^{\pIIL} &= {\cal O}\left(\delta g_{\mu\nu}^{\pII} \right) , \\
		[\delta \ln q]^{\pIIQ} &= {\cal O}\left[ \left(\delta g_{\mu\nu}^{\pI}\right)^2 \right] .
	\end{align}
The first term represents the second-order redshift effects like the Rees-Sciama (RS) effect, 
which are induced by the non-linearities in the Einstein's equations, 
while the second term corresponds to the coupling terms with the lensing, time-delay, and deflection effects in Eqs. (\ref{eq:SGE_PSW}) and (\ref{eq:SGE_PISW}), 
which are induced by the non-linearities in the geodesic equations.
Here, we show the explicit formulae for the first contributions.
The line-of-sight formulae for the second terms are shown in Appendix \ref{sec:LOSc}.
They have forms similar to those for the coupling terms between the two redshift effects, which will be discussed next.

For the second-order contributions, it is sufficient to use the background terms in the source term. 
Hence, the contributions from the first term in Eq. (\ref{eq:qIIsplit}) are given by
	\begin{align}
		\delta I_{{\rm SW}\pII} = \frac{1}{(2\pi)^3}\int_{0}^{\eta_0}\!{\rm d}\eta' \left[ \pd_{\ln q}{\mathfrak S}(\eta') \right]^{\pO} \left[ \int{\rm d}^3k T^{{\rm SW}\pII}({\bs k}, \eta')e^{-i{\bs k} \cdot {\bs n}_{\rm o}(\eta_0-\eta')}  \right] , \label{eq:ISW_II}
	\end{align}
for the SW effect, and
	\begin{align}
		\delta I_{{\rm ISW}\pII} = \frac{1}{(2\pi)^3}\int_{0}^{\eta_0}\!{\rm d}\eta' \left[ \pd_{\ln q}{\mathfrak S}(\eta') \right]^{\pO} \left[ \int{\rm d}^3k \int_{\eta'}^{\eta_0}\!{\rm d}\eta_1 T^{{\rm ISW}\pII}({\bs k}, \eta_1)e^{-i{\bs k} \cdot {\bs n}_{\rm o}(\eta_0-\eta')}  \right] ,\label{eq:IISW_II}
	\end{align}
for the ISW effect (or the RS effect). 
Here,
 	\begin{align}
		T^{{\rm SW}\pII}({\bs k}, \eta') &\equiv -\Psi^{\pII}, \\
		T^{{\rm ISW}\pII}({\bs k}, \eta_1) &\equiv -\left[\Psi^{\pII}-\Phi^{\pII}\right]^{\cdot} + \left[ ik'_j \omega_i^{\pII} + \dot{\chi}_{ij}^{\pII} \right] n_{\rm o}^{(i)}n_{\rm o}^{(j)} .
	\end{align}
Note that, since the background source term is given by,
	\begin{align}
		\left[ \pd_{\ln q}{\mathfrak S} \right]^{\pO} = \bar{g}_v \pd_{\ln q} \bar{I}(q), 
	\end{align}
the integration with respect to $\eta'$ can be analytically performed in Eq. (\ref{eq:IISW_II}) as,
	\begin{align}
		\delta I_{{\rm ISW}\pII} = \frac{\pd_{\ln q} \bar{I}(q)}{(2\pi)^3} \int{\rm d}^3k \int_{0}^{\eta_0}\!{\rm d}\eta_1 e^{-\bar{\tau}(\eta_1)} T^{{\rm ISW}\pII}({\bs k}, \eta_1)e^{-i{\bs k} \cdot {\bs n}_{\rm o}(\eta_0-\eta')}. \label{eq:IISW_II2}
	\end{align}
Here, we have used the facts that the anti-time-ordered integrals in Eq. (\ref{eq:IISW_II}) can be rewritten in terms of the time-ordered ones as, 
	\begin{align}
		\int_{0}^{\eta_0}\!{\rm d}\eta'\int_{\eta'}^{\eta_0}\!{\rm d}\eta_1 = \int_{0}^{\eta_0}\!{\rm d}\eta_1 \int_{0}^{\eta_1}\!{\rm d}\eta' ,
	\end{align}
and that the background intensity is independent of time. \\

- {\it \large Coupling terms between two redshift effects} \\

 Finally, we evaluate the remaining terms, which are quadratic in $\delta \ln q$,
 	\begin{align}
		\delta I_{\rm g}^{\pII} \supset \frac{1}{2}\int_{0}^{\eta_0}\!{\rm d}\eta' \left[\frac{\pd^2 {\mathfrak S}}{\pd \ln q^2}\right]^{\pO}\left([\delta \ln q]^{\pI}\right)^2 .
	\end{align}
It is separated into three terms: SW $\times$ SW, SW $\times$ ISW, and ISW $\times$ ISW terms. 
The first two terms are obtained as,
	\begin{align}
		\delta I_{{\rm SW} \times {\rm SW}} = \frac{\pd_{\ln q}^2 \bar{I}(q)}{2(2\pi)^6}\int{\rm d}^3k_1 \int_{0}^{\eta_0}\!{\rm d}\eta' \bar{g}_v(\eta') T^{{\rm SW}\pI}({\bs k}_1, \eta') \left[ \int{\rm d}^3k_2 T^{{\rm SW}\pI}({\bs k}_2, \eta')e^{-i({\bs k}_1+{\bs k}_2) \cdot {\bs n}_{\rm o}(\eta_0-\eta')}  \right] ,\label{eq:ISWSW}
	\end{align}
and
	\begin{align}
		\delta I_{{\rm SW} \times {\rm ISW}} &= \frac{\pd_{\ln q}^2 \bar{I}(q)}{(2\pi)^6}\int{\rm d}^3k_1 \int_{0}^{\eta_0}\!{\rm d}\eta' \bar{g}_v(\eta') T^{{\rm SW}\pI}({\bs k}_1, \eta') \times \nm \\
		&\qquad \qquad \left[ \int{\rm d}^3k_2 \int_{\eta'}^{\eta_0}\!{\rm d}\eta_1 T^{{\rm ISW}\pI}({\bs k}_2, \eta_1)e^{-i[{\bs k}_1(\eta_0-\eta')+{\bs k}_2(\eta_0-\eta_1)] \cdot {\bs n}_{\rm o}}  \right] . \label{eq:ISWISW}
	\end{align}
These two terms can be absorbed into the intrinsic contributions or the contributions (\ref{eq:IISW_I}) by redefining the source terms appropriately.
This fact is easy to be seen if one performs the expansion only for $[\delta \ln q]_{\rm ISW}$ in Eq. (\ref{eq:mapping2nd}) and uses the momentum that is redshifted through the SW effect, $e^{-D^{\rm SW}}\bar{q}$, in estimating the source terms.
In a similar fashion, 
all contributions with the SW effect can be absorbed into other contributions.

As for the ISW $\times$ ISW terms, we can analytically perform the integration with respect to $\eta'$ as done in Eq. (\ref{eq:IISW_II2}).
Then, it can be written as,
	\begin{align}
		\delta I_{{\rm ISW} \times {\rm ISW}} &= \frac{\pd_{\ln q}^2 \bar{I}(q)}{(2\pi)^6} \int{\rm d}^3k_1 \int_{0}^{\eta_0}\!{\rm d}\eta_2 T^{{\rm ISW}\pI}({\bs k}_1, \eta_2) \times \nm \\
		&\qquad \qquad \left[ \int{\rm d}^3k_2 \int_{0}^{\eta_2}\!{\rm d}\eta_1 e^{-\bar{\tau}(\eta_1)}T^{{\rm ISW}\pI}({\bs k}_2, \eta_1)e^{-i[{\bs k}_1(\eta_0-\eta_2)+{\bs k}_2(\eta_0-\eta_1)] \cdot {\bs n}_{\rm o}}  \right] . \label{eq:IISWISW}
	\end{align}
Note that, here, the domain for the integration with respect to $\eta_1$ is $[0,\eta_2]$. 
However, because of a factor $e^{-\bar{\tau}(\eta_1)}$, 
the integration domain is limited to the free-streaming regime $\eta_{\rm LSS}<\eta_1<\eta_2$.

\subsection{Computation of the bispectrum}\label{ss:bispectrum}

To illustrate how the line-of-sight formulae derived in the previous subsections simplify the calculation, here, 
we apply them to the estimation of the bispectrum, 
	\begin{align}
		B^{m_1m_2m_3}_{l_1l_2l_3}[\delta I]  \equiv \langle a_{l_1m_1}[\delta I] a_{l_2m_2}[\delta I] a_{l_3m_3}[\delta I] \rangle ,
	\end{align}
where 
	\begin{align}
		a_{lm}[\delta I] \equiv \int{\rm d}^2 n_{\rm o} Y_{lm}^{\ast}({\bs n}_{\rm o})\delta I({\bs n}_{\rm o}). 
	\end{align}
Because this subsection is independent of the latter sections, 
those readers primarily interested in the formal aspect of the line-of-sight formulae can skip this subsection in the first reading.

From the discussion in the previous subsection, 
the total second-order perturbations in the intensity are written as,
	\begin{align}
		&\delta I^{\pII} = \int_{0}^{\eta_0}\!{\rm d}\eta' \left\{ \widetilde{\mathfrak S}^{\pII} + \left[\pdif{\widetilde{\mathfrak S}}{\ln q}\right]^{\pO}[\delta \ln q]_{\rm ISW}^{\pIIL} + \left[\pdif{\widetilde{\mathfrak S}}{\ln q}\right]^{\pO}[\delta \ln q]_{\rm ISW}^{\pIIQ} \right. \nm \\ 
		&+ \left. \left[\pdif{\widetilde{\mathfrak S}}{\ln q}\right]^{\pI}[\delta \ln q]_{\rm ISW}^{\pI} +\left[\pdif{\widetilde{\mathfrak S}}{x^i}\right]^{\pI}[\delta x^i]^{\pI} + \left[\pdif{\widetilde{\mathfrak S}}{n^{(i)}}\right]^{\pI}[\delta n^{(i)}]^{\pI} + \frac{1}{2}\left[ \frac{\pd^2 \widetilde{\mathfrak S}}{\pd \ln q\pd \ln q} \right]^{\pO}\left([\delta \ln q]_{\rm ISW}^{\pI}\right)^2 \right\}, \label{eq:mapping2nd_re}
	\end{align}	
where the source term with a tilde is defined by,
	\begin{align}
		\widetilde{\mathfrak S}(x^i, q, n^{(i)}) := {\mathfrak S}(x^i, e^{-D^{\rm SW}(x^i,n^{(i)})}q, n^{(i)}) .
	\end{align}
In Eq. (\ref{eq:mapping2nd_re})	, 
each contribution in the integrand is written as the product of the intrinsic term, which only depends on the perturbed quantities at a position of a non-gravitational scattering source, and the geometrical term, which depend on the metric perturbations between the source and observer. 
Then, as in the standard treatment of CMB lensing, 
the correlations between different combinations of the factors can be treated separately.

The formula (\ref{eq:mapping2nd_re}) also makes it possible to treat the nonlinearities with different origins separately. 
The first two terms arises from nonlinearities in the collision term and Einstein's equations, respectively.
Hence, they are dynamically induced. 
On the other hand, the other four terms have kinematical origins. 
The third term arises because the geodesic equations are nonlinear in the metric perturbations, 
while the four terms in the second line because the intensity changes nonlinearly when the photon geodesics are perturbed.

Now, we show that the contributions to the bispectrum from the kinematically-induced nonlinear terms can be rewritten in a form amenable to numerical calculations. 
The other two dynamically-induced terms have already been considered in Refs. \cite{Huang:2012ub, Su:2012gt, Pettinari:2013he}. 
The kinematically-induced non-linear terms are written as a product of first-order perturbations. 
This type of contributions to the bispectrum was investigated in Ref. \cite{Nitta:2009jp} based on the original line-of-sight formula (\ref{eq:LOSbg}). 
Here, we compute the same type of contributions by using the new line-of-sight formula presented in the previous subsection. 

As an example, we consider the contribution from the scalar mode in the ISW term, the first term in the second line of Eq. (\ref{eq:mapping2nd_re}). 
The other terms are also treated in a similar way with complications in the multipole expansion.
Their explicit expressions will be reported in the subsequent paper with results of their numerical estimation.

Separating the perturbations into the transfer functions and the primordial curvature perturbations $\phi({\bs k})$, 
the ISW term can be written as,
	\begin{align}
		\delta \tilde{I}_{{\rm ISW}\pII} &\equiv \delta I_{{\rm ISW}\pII} + \delta I_{\rm SW \times ISW} \\
		&= \int \frac{{\rm d}^3k_1}{(2\pi)^3} \int \frac{{\rm d}^3k_2}{(2\pi)^3} {\cal T}^{\pII}({\bs k}_1,{\bs k}_2, {\bs n}_{\rm o}) \phi({\bs k}_1)\phi({\bs k}_2) ,
	\end{align}
where
	\begin{align}
		{\cal T}^{\pII}({\bs k}_1,{\bs k}_2, {\bs n}_{\rm o}) &\equiv \int_0^{\eta_0}\!{\rm d}\eta' S(\eta',k_1,\mu_1)e^{-i{\bs k}_1\cdot {\bs n}_{\rm o}(\eta_0-\eta')} \int_{\eta'}^{\eta_0}\!{\rm d}\eta_1 T(\eta_1,k_2)e^{-i{\bs k}_2\cdot {\bs n}_{\rm o}(\eta_0-\eta_1)} ,
	\end{align}
with
	\begin{align}
		S(\eta',k_1,\mu_1) &\equiv \bar{g}_v\left[\pd_{\ln q}I^{\pI}_{00} + \left( i[v_{e}]_{k_1}^{\pI}\mu_1 + \Psi_{k_1}^{\pI}\right)\pd^2_{\ln q} \bar{I} + \pd_{\ln q}\left(I^{\pI}_{20} - \sqrt{6}E^{\pI}_{ 20}\right)\frac{P_2(\mu_1)}{10}\right], \label{eq:TransferS}\\
		T(\eta_1,k_2) &\equiv -\left[\Psi_{k_2}^{\pI}-\Phi_{k_2}^{\pI}\right]^{\cdot}, \label{eq:TransferT}
	\end{align}
and $\mu_1 \equiv {\bs n}_{\rm o} \cdot \hat{\bs k}_1$.
$P_2(\mu_1)$ is the second Legendre polynomial.
In deriving Eq. (\ref{eq:TransferS}), 
we have introduced the velocity potential through $[{\bs v}_e]_{k_1} \equiv i\hat{\bs k}_1[v_e]_{k_1}$ and the distribution function for the E-mode polarization, $E$ \cite{Beneke:2010eg}. 
$I_{00}$, $I_{20}$, and $E_{20}$ represent the multipole moments for the corresponding quantities with the $z$-direction parallel to $\hat{\bs k}_1$.
Note that the dependence on $\mu_1$ (and then ${\bs n}_{\rm o}$ and $\hat{\bs k}_1$) can be eliminated from the function $S$ by replacing it with the time derivative as,
	\begin{align}\label{eq:trick}
		S(\eta',k_1,\mu_1)e^{-i{\bs k}_1\cdot {\bs n}_{\rm o}(\eta_0-\eta')} = S\left(\eta',k_1,\frac{1}{ik_1}\dif{}{\eta'}\right)e^{-i{\bs k}_1\cdot {\bs n}_{\rm o}(\eta_0-\eta')} .
	\end{align}
	
Then, the multipole moments are given by,
	\begin{align}
		a_{lm}[\delta \tilde{I}_{{\rm ISW}\pII}] = (4\pi)^2 (-i)^l\int \frac{{\rm d}^3k_1}{(2\pi)^3} \int \frac{{\rm d}^3k_2}{(2\pi)^3} {\cal T}^{\pII}_{lm}({\bs k}_1,{\bs k}_2) \phi({\bs k}_1)\phi({\bs k}_2) , \label{eq:Ilm}
	\end{align}
where
	\begin{align}
		{\cal T}^{\pII}_{lm}({\bs k}_1,{\bs k}_2) &=
		\sum_{l_1,m_1,l_2,m_2}i^{l_1+l_2+l}{\cal G}^{m_1m_2m}_{l_1l_2l}
		\int_0^{\eta_0}\!{\rm d}\eta' Y_{l_1m_1}(-\hat{\bs k}_1) S\left(\eta',k_1,\frac{1}{ik_1}\dif{}{\eta'}\right) j_{l_1}\left[k_1(\eta_0-\eta')\right] \nm \\
		&\qquad \qquad \times \int\!{\rm d}\eta_1 Y_{l_2m_2}(-\hat{\bs k}_2) T(k_2, \eta_1) j_{l_2}\left[k_2(\eta_0-\eta_1)\right] , \label{eq:tlm}
	\end{align}
using the Gaunt integral, 
	\begin{align}
		{\cal G}^{m_1m_2m}_{l_1l_2l} \equiv \int {\rm d}^2n_{\rm o} Y_{l_1m_1}({\bs n}_{\rm o})Y_{l_2m_2}({\bs n}_{\rm o})Y_{lm}({\bs n}_{\rm o}) .
	\end{align}
On the other hand, at first order, 
the intensity can be written in the form, 
	\begin{align}
		a_{lm}[\delta I^{\pI}] = 4\pi (-i)^l\int\frac{{\rm d}^3k}{(2\pi)^3} Y^{\ast}_{lm}(\hat{\bs k}){\cal T}_l^{\pI}(k)\phi({\bs k}) .
	\end{align}
Thus, the bispectrum can be written in the form,
	\begin{align}\label{eq:LOS_bispectrum}
		B^{m_1m_2m_3}_{l_1l_2l_3}[\delta I] &= 2 {\cal G}^{m_1m_2m_3}_{l_1l_2l_3} \int_0^{\eta_0}\!{\rm d}\eta' b^S_{l_1}(\eta') b^T_{l_2}(\eta') + \text{2 sym.},
	\end{align}
where
	\begin{align}
		b^S_{l_1}(\eta') &\equiv \frac{2}{\pi}\int k_1^2 {\rm d}k_1 P_{\phi}(k_1){\cal T}_{l_1}^{(I)}(k_1)S\left(\eta',k_1,\frac{1}{ik_1}\dif{}{\eta'}\right) j_{l_1}\left[k_1(\eta_0-\eta')\right] , \label{def:bs}
	\end{align}
and
	\begin{align}
		b^T_{l_2}(\eta') &\equiv \frac{2}{\pi} \int_{\eta'}^{\eta_0}\!{\rm d}\eta_1 \int k_2^2{\rm d}k_2 P_{\phi}(k_2){\cal T}_{l_2}^{(I)}(k_2)T(k_2, \eta_1)j_{l_2}\left[k_2(\eta_0-\eta_1)\right] , \label{def:bt}
	\end{align}
with $P_{\phi}$ being the primordial power spectrum.
The infinite sums over multipoles in Eq. (\ref{eq:tlm}) has disappeared in Eq. (\ref{eq:LOS_bispectrum}) because of the orthogonality of the spherical harmonics $Y_{lm}(\hat{\bs k})$. 
\footnote{Note that this result is a consequence of the statistical homogeneity and isotropy of the perturbations at positions of sources.}
As also mentioned in Ref. \cite{Nitta:2009jp}, the bispectrum from products of the first-order perturbations can be written in a form similar to the standard formula for the local-type bispectrum, 
where the integration over the Fourier modes are split into one-dimensional integrations \cite{Komatsu:2001rj}.
Here, in addition, 
no infinite sum over multipoles appears in the expression (\ref{eq:LOS_bispectrum}) 
because the source terms are now written in terms of a finite number of the multipole moments of the intensity (or brightness).
This is also the case for the other kinematically-induced nonlinear terms. 

Using the first-order line-of-sight formula (\ref{eq:LOS1st}), 
the first-order transfer function ${\cal T}^{\pI}_l$ can be written as an integral over the product of the source term and the spherical Bessel function.
Because the factors other than the spherical Bessel functions slowly vary with respect to $k_i~(i=1,2)$ in the integrand of $b^S_{l_1}$ and $b^T_{l_2}$, 
we will be able to apply the Limber approximation \cite{1953ApJ...117..134L} (see also Ref. \cite{Lewis:2006fu}) in their estimation.
 Note that the similar approximation cannot be applied 
when the source term includes the contribution from the gravitational collision terms.
In this case, the function corresponding to $S$ depends on the intensity in the free-streaming regime, 
which rapidly varies with respect to the wavenumber.

It is also noteworthy that $b^T_{l_2}(\eta')$ varies slowly with respect to $\eta'$.
Then, its $\eta'$-dependence can be neglected if a source is sufficiently thin in the radial direction. 
For example, the width of the last scattering surface to the entire integration range is of order $0.01$.
Once its width is neglected, the contribution to the bispectrum from the last scattering surface is factorized into a product of the intrinsic and geometric terms. 
These two terms, which depend respectively on phenomena before and after the photon decoupling, 
can be separately estimated.

\section{Curve-of-sight formulae: moments}\label{sec:LOS_m}

\subsection{Brightness}\label{ss:LOS_b}

 Since we have obtained the line-of-sight formulae for the intensity $I$, 
 it is straightforward to obtain formulae for the fractional brightness by taking its third moment:
 	\begin{align}\label{def:delta2}
		\Delta \equiv \frac{\delta B}{\bar{B}}; \quad B \equiv \int\!{\rm d}q_0~ q_0^3 I.
	\end{align}
 
 Since only the source terms depend on $q_0$ and the line-of-sight formulae presented in the previous section are linear in it, 
 the line-of-sight formulae for the brightness are obtained from those for the intensity by replacing the source terms as, 
 	\begin{align}
		k^i{\mathfrak S} &\to k^i{\mathfrak S}^{\Delta}, \\
		\pd_{n^{(i)}}{\mathfrak S} &\to \pd_{n^{(i)}}{\mathfrak S}^{\Delta},
	\end{align}
for the lensing, time-delay, and deflection effects, where
	\begin{align}
		{\mathfrak S}^{\Delta} &\equiv \frac{1}{\bar{B}}\int\!{\rm d}q_0~ q_0^3 {\mathfrak S}.
	\end{align} 
As for the source terms for the redshift effects, 
they should be replaced as,
	\begin{align}
		\pd_{\ln q}{\mathfrak S} \to -4{\mathfrak S}^{\Delta} ,
	\end{align}
and 
	\begin{align}
		\pd^2_{\ln q}{\mathfrak S} \to 16{\mathfrak S}^{\Delta} .
	\end{align}
 To check its correctness, 
 let us see the first-order contributions. 
 At first order, only the linear redshift effects contribute.
They are evaluated as,
	\begin{align}
		\Delta_{{\rm SW}\pI} 
		= -4\int_{0}^{\eta_0}\!{\rm d}\eta' \bar{g}_v(\eta') \left[ \int \frac{{\rm d}^3k'}{(2\pi)^3} T^{{\rm SW}\pI}({\bs k}', \eta')e^{-i{\bs k}' \cdot {\bs n}_{\rm o}(\eta_0-\eta')}  \right] ,
	\end{align}
for the SW effect, and
	\begin{align}
		\Delta_{{\rm ISW}\pI} &= -4\int_{0}^{\eta_0}\!{\rm d}\eta' \bar{g}_v(\eta') \left[ \int \frac{{\rm d}^3k'}{(2\pi)^3} \int_{\eta'}^{\eta_0}\!{\rm d}\eta_1 T^{{\rm ISW}\pI}({\bs k}', \eta_1)e^{-i{\bs k}' \cdot {\bs n}_{\rm o} (\eta_0-\eta_1)}  \right] \\
		&=  -4\int \frac{{\rm d}^3k'}{(2\pi)^3} \int_{0}^{\eta_0}\!{\rm d}\eta_1 e^{-\bar{\tau}(\eta_1)}T^{{\rm ISW}\pI}({\bs k}', \eta_1)e^{-i{\bs k}' \cdot {\bs n}_{\rm o} (\eta_0-\eta_1)} ,
	\end{align}
for the ISW effect. 
Then, the gravitational contribution at first order becomes
	\begin{align}
		\Delta_G^{\pI} = 4\int \frac{{\rm d}^3k'}{(2\pi)^3}  \int_{0}^{\eta_0}\!{\rm d}\eta \left\{ \bar{g}_v(\eta) \Psi^{\pI} + e^{-\bar{\tau}(\eta)}\left[(\Psi^{\pI}-\Phi^{\pI})^{\cdot} - \dot{\chi}^{\pI}_{ij}n^{(i)}_{\rm o} n^{(j)}_{\rm o}\right] \right\}e^{-i{\bs k}' \cdot {\bs n}_{\rm o} (\eta_0-\eta)} ,
	\end{align}
or, using the relation between the fractional brightness and the temperature fluctuations $\Delta^{\pI}=4\Theta^{\pI}$, 
	\begin{align}
		\Theta_G^{\pI} &= \int \frac{{\rm d}^3k'}{(2\pi)^3}  \int_{0}^{\eta_0}\!{\rm d}\eta \left\{ \bar{g}_v(\eta) \Psi^{\pI} + e^{-\bar{\tau}(\eta)}\left[(\Psi^{\pI}-\Phi^{\pI})^{\cdot} - \dot{\chi}^{\pI}_{ij}n^{(i)}_{\rm o} n^{(j)}_{\rm o}\right] \right\}e^{-i{\bs k}' \cdot {\bs n}_{\rm o} (\eta_0-\eta)}. \label{eq:1stT}
	\end{align}
Thus, 
our result correctly reproduces the first-order terms dependent on the metric perturbations. 

\subsection{Derivation from the brightness equation}\label{ss:derivation_b}
  As expected, the same line-of-sight formulae are obtained directly from the brightness equation (\ref{eq:brightness}). 
  The brightness equation can be rewritten as,
  	\begin{align}
		\dif{}{\eta}\left[e^{-4\delta \ln q} (1+\Delta) \right] = e^{-4\delta \ln q}{\mathfrak C}^{\Delta}[\Delta] ,
	\end{align}
 where
 	\begin{align}
		\delta \ln q(\eta) \equiv -D^{\rm SW}[\eta, x^i(\eta),n^{(i)}(\eta)] - \int_{\eta}^{\eta_0} {\rm d}\eta_1 D^{\rm ISW}[\eta_1,x^i(\eta_1),n^{(i)}(\eta_1)] ,
	\end{align}
with $x^i(\eta)$ and $n^{(i)}(\eta)$ being solutions of the full geodesic equations.
Repeating the same manipulations for the Boltzmann equation, 
the gravitational contributions can be written as,
	\begin{align}
		\Delta_G(\eta_0, x_0^i,n_0^{(i)}) = \int_{0}^{\eta_0}\!{\rm d}\eta' \left\{ e^{-4\delta \ln q(\eta')}{\mathfrak S}^{\Delta}[\Delta; \eta', x^i(\eta),n^{(i)}(\eta)] - {\mathfrak S}^{\Delta}[\Delta; \eta', \bar{x}^i(\eta),\bar{n}^{(i)}(\eta)] \right\}. \label{eq:mapping_delta}
	\end{align}
Expanding this equation with respect to the deviations $\delta \ln q$, $\delta x^i$, and $\delta n^{(i)}$, 
we get the line-of-sight formulae for the brightness obtained in the previous subsection.

\subsection{General moments and spectral distortions}\label{ss:LOS_sd}
It is also straightforward to obtain line-of-sight formulae for a general moment of the intensity,
	\begin{align}
		{\cal M}_n[I] := \int\!{\rm d}q~ q^n I .
	\end{align}
For the $n$-th moment, its gravitational contributions are obtained by replacing the source term in the line-of-sight formulae for the intensity as,
	\begin{align}
		k^i{\mathfrak S} &\to k^i{\cal M}_n[{\mathfrak S}], \\
		\pd_{n^{(i)}}{\mathfrak S} &\to \pd_{n^{(i)}}{\cal M}_n[{\mathfrak S}],
	\end{align}
for the lensing, time-delay, and deflection effects. 
As for the redshift effects,
	\begin{align}
		\pd_{\ln q}{\mathfrak S} \to -(n+1){\cal M}_n[{\mathfrak S}] ,
	\end{align}
and 
	\begin{align}
		\pd^2_{\ln q}{\mathfrak S} \to (n+1)^2{\cal M}_n[{\mathfrak S}] .
	\end{align}
Since these moments characterize the $q$-dependence of the intensity completely, 
their line-of-sight formulae are expected to give those of the spectral distortions. 
For example, 
the $y$ distortion can be evaluated through the second and third moments. 
In fact, 
it can be written in terms of these two moments as,
	\begin{align}
		\ln(1+4y) = \ln\left(\frac{{\cal M}_3[I]}{{\cal M}_3[\overline{I}]}\right) - \frac{4}{3}\ln\left(\frac{{\cal M}_2[I]}{{\cal M}_2[\overline{I}]}\right) ,
	\end{align}
by using the relation ${\cal M}_3[I] \propto T^4(1+4y)$ and ${\cal M}_2[I] \propto T^3$. 

\section{Remapping: relation to the standard treatment of CMB lensing}\label{sec:remapping}

Finally, 
we discuss the relation between our approach and the standard treatment of CMB lensing as remapping of the apparent directions.
 See also Ref. \cite{Su:2014mga} for another approach to derive the remapping formula from the Boltzmann equation.
 
\begin{figure}[htbp] 
		\centering
		\includegraphics[width=.8\linewidth]{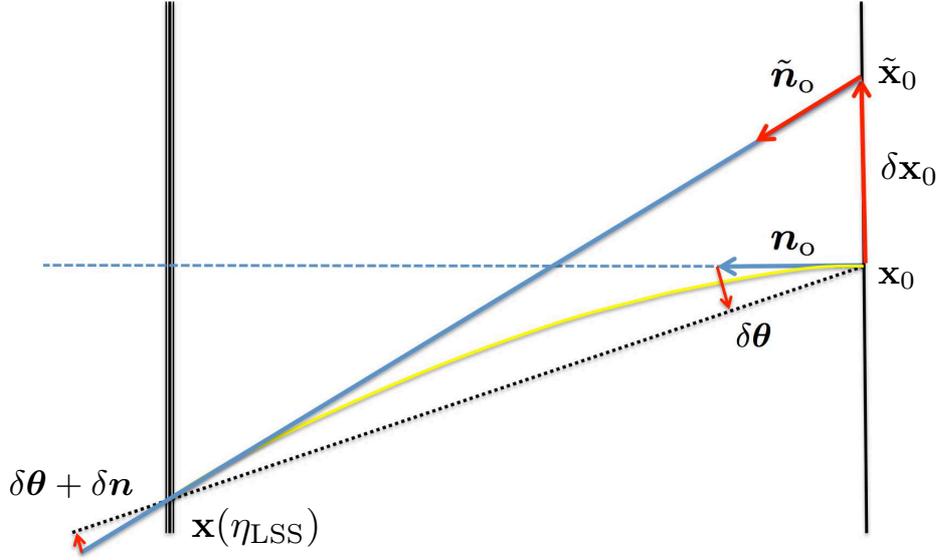}
		\caption{Remapping of the coordinates. The straight dashed line and curved solid line represent a background geodesic and a full geodesic, respectively. The remapped coordinates of the observation $(\tilde{\eta}_0, \tilde{z}_0^{(A)})$ are obtained by pulling back the coordinates in the last scattering surface $(\eta_{\rm LSS}, z^{(A)}(\eta_{\rm LSS}))$ along a background geodesic represented by the straight solid line.}
		\label{fig:remapping}
	\end{figure}

Here, 
we work under the thin last-scattering-surface approximation and also neglect non-gravitational scatterings after the last scattering. 
As mentioned in the subsection \ref{ss:difficulty}, the intensity $I$ is conserved along a perturbed geodesic. 
Correspondingly, the intensity without the gravitational effects $I_{\rm LSS}$, which is defined by Eq. (\ref{def:ILSS}), is conserved along a background geodesic. 
Then, by pulling back the coordinates in the last scattering surface along a background geodesic (see Fig. \ref{fig:remapping}), 
the full intensity $I$ can be written in terms of $I_{\rm LSS}$ as,
	\begin{align}\label{eq:IILSS}
		I(\eta_0, z_0^{(A)}) = I_{\rm LSS}(\tilde{\eta}_0, \tilde{z}_0^{(A)}) .
	\end{align}
Here, the remapped coordinates of the observation $(\tilde{\eta}_0, \tilde{z}_0^{(A)})$ are determined through the relations,
	\begin{align}
		\bar{x}^i(\eta_{\rm LSS};\tilde{\eta}_0, \tilde{z}_0^{(A)}) &= x^i(\eta_{\rm LSS};\eta_0, z_0^{(A)}) , \label{eq:remap_x} \\
		\bar{q}(\eta_{\rm LSS};\tilde{\eta}_0, \tilde{z}_0^{(A)}) &= q(\eta_{\rm LSS};\eta_0, z_0^{(A)}), \label{eq:remap_q} \\
		\bar{n}^{(i)}(\eta_{\rm LSS};\tilde{\eta}_0, \tilde{z}_0^{(A)}) &= n^{(i)}(\eta_{\rm LSS};\eta_0, z_0^{(A)}) .\label{eq:remap_n}
	\end{align}
Thus, the perturbations $\delta I_G$ is now written in terms of the remapped coordinates as,
	\begin{align}\label{eq:remapping_I}
		\delta I_{\rm G}(\eta_0, z_0^{(A)}) = I_{\rm LSS}(\tilde{\eta}_0, \tilde{z}_0^{(A)}) - I_{\rm LSS}(\eta_0, z_0^{(A)}),
	\end{align}
as in the standard treatment of CMB lensing. 
One can take into account the width of the last scattering surface by considering remapping for each time slice in the last scattering epoch.
In a similar fashion, 
it is also possible to include the contributions from the ISW term, 
whose support extends over the free-streaming regime.

The corresponding equation for the brightness can be obtained by taking the third moment with respect to $q_0$.
Using the solution for the geodesic equations (\ref{eq:SGE_Pq}), 
the equation (\ref{eq:remap_q}) can be solved for $\tilde{q}_0$ as,
	\begin{align}
		\tilde{q}_0 = e^{\delta \ln q(\eta_{\rm LSS})}q_0.
	\end{align}
Note that the solution $\delta \ln q(\eta)$ does not depend on $q_0$.
Therefore, 
the remapping formula for the fractional brightness is obtained as,
	\begin{align}\label{eq:remapping_b}
		\Delta_{\rm G}(\eta_0, x_0^{(i)},n_0^{(i)}) = e^{-4\delta \ln q(\eta_{\rm LSS})}\Delta_{\rm LSS}(\tilde{\eta}_0, \tilde{x}_0^i, \tilde{n}_0^{(i)}) - \Delta_{\rm LSS}(\eta_0, x_0^{(i)},n_0^{(i)}) ,
	\end{align}
where $\Delta_{\rm LSS}$ is defined by,
	\begin{align}
		\Delta_{\rm LSS} \equiv \frac{\delta B_{\rm LSS}}{\bar{B}}; \quad B_{\rm LSS} \equiv \int\!{\rm d}q_0~ q_0^3 I_{\rm LSS}.
	\end{align}
	
Hereafter, to see the relation to the standard treatment of CMB lensing, 
we concentrate on changes in the brightness due to the deviations in a photon trajectory and separate the redshift effect $e^{-4\delta \ln q(\eta_{\rm LSS})}$:
	\begin{align}\label{eq:remapping_b2}
		\Delta_{\rm g}(\eta_0, x_0^{(i)},n_0^{(i)}) \equiv \Delta_{\rm LSS}(\tilde{\eta}_0, \tilde{x}_0^i, \tilde{n}_0^{(i)}) - \Delta_{\rm LSS}(\eta_0, x_0^{(i)},n_0^{(i)}) .
	\end{align}
In terms of the deviations, the relations (\ref{eq:remap_x}) and (\ref{eq:remap_n}) are given by,
	\begin{align}
		\delta x_{0\parallel} &\equiv n_{{\rm o}(i)}(\tilde{x}_0^i - x_0^i) = -\delta \eta_0 + \delta x_{\parallel}(\eta_{\rm LSS}) , \label{eq:remap_dparallel} \\
		\delta x_{0\perp}^i &\equiv S_{\rm o}^{(i)(j)}(\tilde{x}_{0i}- x_{0i}) = S_{\rm o}^{(i)(j)}\delta n_{0(j)}(\tilde{\eta}_0 - \eta_{\rm LSS}) + \delta \theta^{(i)}(\eta_0-\eta_{\rm LSS}), \label{eq:remap_dperp} \\
		\delta n_0^{(i)} & \equiv \tilde{n}_0^{(i)} - n_0^{(i)} = \delta n^{(i)}(\eta_{\rm LSS}) \label{eq:remap_dn},
	\end{align}
with $\delta \eta_0 \equiv \tilde{\eta}_0 -\eta_0$. 
Here, we have used the fact that the perpendicular components $\delta x_{\perp}^i(\eta_{\rm LSS})$ can be written in terms of the change in the observed angle $\delta \theta^{(i)}$ as $\delta x_{\perp}^i(\eta_{\rm LSS}) = \delta \theta^{(i)}(\eta_0-\eta_{\rm LSS})$.
Since the time $\tilde{\eta}_0$ is arbitrary, 
we can choose its value so that the parallel component $\delta x_{0\parallel}$ vanishes:
	\begin{align}
		\delta \eta_0 = \delta x_{\parallel}(\eta_{\rm LSS}). 
	\end{align}
However, we cannot set $\delta x_{0\perp}^i = 0$ in general as assumed in the standard treatment.
Then, we need to make an approximation here. 
The naive approximation
	\begin{align}\label{eq:naive_approx}
		\Delta_{\rm LSS}(\tilde{\eta}_0, \tilde{x}_0^i, \tilde{n}_0^{(i)}) \simeq \Delta_{\rm LSS}(\tilde{\eta}_0, x_0^i, \tilde{n}_0^{(i)}) ,
	\end{align}
is not so accurate because of the large factor $\eta_0-\eta_{\rm LSS}$ in Eq. (\ref{eq:remap_dperp}). 
In fact, by using the explicit formulae in the subsection \ref{ss:deviations}, $\delta x_{0\perp}^i$ is roughly estimated to be,
	\begin{align}
		\delta x_{0\perp}^i = k_2(\eta_0-\eta_{\rm LSS})^2{\cal O}(|\delta g_{\mu\nu}|) ,
	\end{align}
for the Fourier modes of the metric perturbations with a wavenumber $k_2$. 
Here, ${\cal O}(|\delta g_{\mu\nu}|)$ represents terms of the order of the metric perturbations $\Psi$ and $\Phi$.

Then, the error in the approximation (\ref{eq:naive_approx}) is 
	\begin{align}
		\Delta_{\rm LSS}(\tilde{\eta}_0, x_0^i, \tilde{n}_0^{(i)}) = \left[ 1 + k_1k_2(\eta_0-\eta_{\rm LSS})^2{\cal O}(|\delta g_{\mu\nu}|) \right] \Delta_{\rm LSS}(\tilde{\eta}_0, \tilde{x}_0^i, \tilde{n}_0^{(i)}) ,
	\end{align}
where $k_1$ is a wavenumber of $\Delta_{\rm LSS}$.
Since $k_i(\eta_0-\eta_{\rm LSS})~(i=1,2)$ is of the order of the observed multipoles, the error can be non-negligibe.

Instead, 
one can make the approximation more accurate by also changing the direction $\tilde{n}_0^{(i)}$ as,
	\begin{align}\label{eq:improved_approx}
		\Delta_{\rm LSS}(\tilde{\eta}_0, \tilde{x}_0^i, q_0, \tilde{n}_0^{(i)}) \simeq \Delta_{\rm LSS}(\tilde{\eta}_0, x_0^i, q_0, \tilde{\tilde{n}}_0^{(i)}) ,
	\end{align}
where
	\begin{align}
		\tilde{\tilde{n}}_0^{(i)} \equiv n_0^{(i)} - \frac{\eta_{\rm LSS}-\eta_0}{\eta_{\rm LSS}-\tilde{\eta}_0}\delta \theta^{(i)} .
	\end{align}
This equation can be rewritten in terms of the observed direction $n_{\rm o}^i \equiv -n_0^{(i)}$ as,
	\begin{align}
		\tilde{\tilde{n}}_{\rm o}^{(i)} = n_{\rm o}^{(i)} + \frac{\eta_{\rm LSS}-\eta_0}{\eta_{\rm LSS}-\tilde{\eta}_0}\delta \theta^{(i)} .
	\end{align}  
Here, $\tilde{\tilde{n}}_0^{(i)}$ has been chosen so that $\delta \tilde{\tilde{n}}_0^{(i)} \equiv \tilde{\tilde{n}}_0^{(i)} - n_0^{(i)}$ solves the equation $\delta x_{0\perp}^i = 0$. 
From Eq. (\ref{eq:parallel}), the prefactor of $\delta \theta^{(i)}$ is estimated to be
	\begin{align}
		\frac{\eta_{\rm LSS}-\eta_0}{\eta_{\rm LSS}-\tilde{\eta}_0} = 1 + {\cal O}(|\delta g_{\mu\nu}|) .
	\end{align}
Then, it is a good approximation to set
	\begin{align}
		\tilde{\tilde{n}}_{\rm o}^{(i)} \simeq n_{\rm o}^{(i)} + \delta \theta^{(i)} .
	\end{align}
From the definition, the LHS of Eq. (\ref{eq:improved_approx}) can be written as,
	\begin{align}
		\Delta_{\rm LSS}(\eta_0, \tilde{x}_0^i, \tilde{n}_0^{(i)}) &\simeq S^{\Delta}[\Delta;\eta_{\rm LSS}, x^i(\eta_{\rm LSS}), n^{(i)}(\eta_{\rm LSS})] ,
	\end{align}
in the thin last-scattering-surface approximation.
On the other hand, the RHS is
	\begin{align}
		\Delta_{\rm LSS}(\tilde{\eta}_0, x_0^i, \tilde{\tilde{n}}_0^{(i)}) &\simeq  S^{\Delta}[\Delta;\eta_{\rm LSS}, x^i(\eta_{\rm LSS}), n_0^{(i)} - \delta \theta^{(i)}] , \nm \\
		&= S^{\Delta}[\Delta;\eta_{\rm LSS}, x^i(\eta_{\rm LSS}), n^{(i)}(\eta_{\rm LSS}) - \delta n^{(i)} - \delta \theta^{(i)}] .
	\end{align}
Thus, the error in the approximation (\ref{eq:improved_approx}) is estimated to be,
	\begin{align}
		\Delta_{\rm LSS}(\tilde{\eta}_0, x_0^i, \tilde{\tilde{n}}_0^{(i)}) =  \left[ 1+  l_1 k_2(\eta_0-\eta_{\rm LSS}){\cal O}(|\delta g_{\mu\nu}|) \right]\Delta_{\rm LSS}(\tilde{\eta}_0, \tilde{x}_0^i, \tilde{n}_0^{(i)}),
	\end{align}
where $l_1$ is a multipole of the source term $S^{\Delta}$, which is of the order of unity.
Then, the error is reduced to the acceptable level.
Making the approximation (\ref{eq:improved_approx}), 
the remapping formula (\ref{eq:remapping_b2}) is now written in the form used in the standard treatment of CMB lensing,
	\begin{align}
		\Delta_{\rm g}(\eta_0, x_0^i, n_0^{(i)}) \simeq \Delta_{\rm LSS}(\eta_0 + \delta \eta_0, x_0^i, n_0^{(i)} - \delta \theta^{(i)}) - \Delta_{\rm LSS}(\eta_0, x_0^i,n_0^{(i)}) .
	\end{align}
From Eqs. (\ref{eq:parallel}) and (\ref{eq:angle}), one can also find that the expressions for time delay $\delta \eta_0$ and the change in the observed angle $\delta \theta^{(i)}$ coincide with the standard ones.
Note that the time-delay effect is of the same order of magnitude as the neglected effects. 
Then, it should also be neglected as done in the standard treatment.

\section{Summary}\label{sec:summary}
In this paper, 
we introduced a new approach to the treatment of the gravitational effects such as redshift, time delay, and lensing on the observed CMB anisotropies based on the Boltzmann equation.
Motivated by the Liouville's theorem in curved spacetime, 
we first derived the mapping formula that relates the observed intensity to the non-gravitational scattering sources.
This formula can be considered as a generalization of the remapping formula in the standard treatment of CMB lensing.
In a similar fashion that the lensing effect is treated as remapping of the apparent directions in the standard treatment, 
in this formula, all the gravitational effects appear as changes in the mapping between the observer and sources. 
Here, the redshift, time delay, and lensing effects are treated on the same footing. 
In addition to these effects, the formula also includes the contributions from deflection at a position of a source.
Introducing the remapped coordinates, 
we also showed that the remapping formula in the standard treatment is reproduced when the subleasing effects are neglected.
In this approach, the geometrical meanings of the approximations are easy to be understood.

Next, we derived a second-order line-of-sight formula for each gravitational effect by expanding the mapping formula up to second order.
In this approach, the separation of the gravitational and intrinsic effects are manifest and guaranteed automatically.
This formula clarifies the multipole dependence of the observed intensity and give an efficient way to estimate its bispectrum.

This approach to the treatment of the gravitational effects provides a tool to analyze the entire evolution of the CMB anisotropies in a uniform way without spoiling the geometrical intuition in the remapping approach of CMB lensing.
Because the Boltzmann equation for polarization has a similar structure, 
the extension to polarization will also be easy in this approach. 
The full treatment including polarization will be presented in the upcoming paper. 

\vspace{24pt}

{\bf Note added:} Another group has also developed a similar approach in Ref. \cite{Fidler:2014zwa}. They also introduce a new approach to treat all gravitational effects based on the Boltzmann equation, where the gravitational effects are treated as a transformation operator of the distribution function, ${\cal J}(\eta)$. Our line-of-sight formulae in Sec. \ref{sec:LOS_df} are reproduced in their approach when ${\cal J}(\eta)$ is solved backwards in time with ${\cal J}(\eta_0)=1$. On the other hand, their line-of-sight formulae are derived for ${\cal J}(\eta)$ that is solved forward in time. They are more related to ours in Sec. \ref{sec:remapping}. Under the same approximations in Sec. \ref{sec:remapping}, they coincide with those derived from the remapping formula (\ref{eq:remapping_I}). Here, the transformation operator ${\cal J}(\eta_0)$ corresponds to the derivative operator that appears after the expansion of the RHS of Eq. (\ref{eq:IILSS}) with respect to $\delta z_0^{(A)} (\equiv \tilde{z}_0^{(A)} - z_0^{(A)})$.

\vspace{24pt}

{\bf Acknowledgments}
\vspace{10pt}

We thank C. Fidler and K. Koyama for reading the manuscript and sharing information before the submission. 
We also thank Z. Huang and F. Vernizzi for reading the manuscript and helping us to understand their works correctly.
R.~S. is supported by Grant-in-Aid for JSPS postdoctoral fellowships for research abroad. 
A.~N is grateful to YITP for the hospitality during his stay when part 
of this work was done. 
A.~N. is supported by Grant-in-Aid for JSPS Fellows No. 26-3409. 
T.~H. is supported in part by MEXT SPIRE and JICFuS.
This work is supported by the JSPS Grant-in-Aid for Scientific 
Research (A) No. 21244033. 

\appendix

\section{Relation of our definition of the metric to the standard one}\label{sec:metric_def}
In this paper, 
we used the metric in the ADM form:
	\begin{align}\label{def:metric_app}
		{\rm d}s^2 = a(\eta)^2\left[-e^{2\Psi}{\rm d}\eta^2 + \gamma_{ij}({\rm d}x^i + \omega^i{\rm d}\eta)({\rm d}x^j + \omega^j{\rm d}\eta) \right],
	\end{align}
where
	\begin{align}
		[\ln {\bs \gamma}]_{ij} \equiv 2h_{ij} \equiv 2\Phi\delta_{ij} + 2\chi_{ij}, \label{def:tensor_app}
	\end{align}
with ${\omega^{i}}_{,i}=0$ and $\chi^i_i={\chi_i^j}_{,j}=0$. 
Here, 
we show how our metric perturbations $\Psi$, $\Phi$, $\omega^i$, and $\chi_{ij}$ are related to those in the standard definition in the conformal Poisson gauge,
	\begin{align}\label{def:metric_std}
		{\rm d}s^2 = a(\eta)^2\left[-(1+2A){\rm d}\eta^2 + 2B_{i}{\rm d}x^i{\rm d}\eta + ((1+2D)\delta_{ij} + 2E_{ij}){\rm d}x^i {\rm d}x^j \right],
	\end{align}
where $B^i_{,i}=0$ and $E^i_i = E^i_{j,i}=0$.
Note that the indices for the perturbations are lowered and raised with the Kronecker's delta.  

In terms of $\Psi$, $\Phi$, $\omega^i$, and $\chi_{ij}$, 
the $00$- and $0i$-components of the metric can be written as
	\begin{align}
		g_{00} &= -e^{2\Psi} + \gamma_{ij}\omega^i\omega^j \simeq -(1+ 2\Psi + 2\Psi^2) , \\
		g_{0i} &= a^2\gamma_{ij}\omega^i \simeq a^2\delta_{ij}\omega^i, 
	\end{align}
to second order. 
The trace and traceless parts of the spatial metric $\gamma_{ij}$ can be read from Eq. (\ref{eq:expand_gamma}) as
	\begin{align}
		{\rm tr}({\bs \gamma}) &\simeq e^{2\Phi}(3+2\chi^{kl}\chi_{kl}) ,\\
		\hat{\gamma}_{ij} &\equiv \gamma_{ij} - \frac{{\rm tr}({\bs \gamma})}{3}\delta_{ij} \simeq 2e^{2\Phi}\left(\chi_{ij} + \chi_i^k \chi_{kj} - \frac{\chi^{kl}\chi_{kl}}{3}\delta_{ij} \right) .
	\end{align}
The $0i$-components satisfy the gauge condition $\pd^i g_{0i}=0$, 
but the traceless part of the spatial metric does not, $\pd^ i \hat{\gamma}_{ij} \neq 0$.
Hence, 
we need to make a gauge transformation to find the correspondence with $A$, $B_i$, $D$, and $E_{ij}$.
Since $\pd^ i \hat{\gamma}_{ij}$ is a second-order quantity, 
	\begin{align}
		\pd^ i \hat{\gamma}_{ij} = 4\chi_{ij}\pd^i \Phi + 2\pd^i\left( \chi_i^k \chi_{kj} - \frac{\chi^{kl}\chi_{kl}}{3}\delta_{ij} \right) ,
	\end{align}
it is sufficient to consider a gauge transformation associated to
	\begin{align}
		\tilde{x}^{\mu} &= x^{\mu} + \xi^{\mu}(x^{\mu}) ,
	\end{align}
where $\xi^i$ is perturbatively expanded as
	\begin{align}
		\xi^{\mu} = [\xi^{\mu}]^{(II)} + \cdots.
	\end{align}
Decomposing $\xi^i$ into scalar and vector modes as
	\begin{align}
		\xi^i &= \xi^{,i} + \xi^i_{\perp} , 
	\end{align}
with $\pd_i \xi^i_{\perp}=0$, it is required that they should satisfy
	\begin{align}
		\xi^0 &= -\dot{\xi}, \\
		\Delta^2 \xi &= \frac{3}{4}\pd^k\pd^l \hat{\gamma}_{kl}, \label{eq:xi} \\
		\Delta \xi_{\perp i} &= \left( \delta^l_i - \frac{\pd_i\pd^l}{\Delta} \right)\pd^k \hat{\gamma}_{kl}, \label{eq:xi_perp}
	\end{align}
in order that the metric satisfies the gauge conditions $\pd^i g_{0i}=0$ and $\pd^ i \hat{\gamma}_{ij} = 0$ after the gauge transformation. 
 Hence, 
 provided that $\xi$ and $\xi_{\perp}^i$ are the solutions of Eqs. (\ref{eq:xi}) and (\ref{eq:xi_perp}), 
 $A$, $B_i$, $D$, and $E_{ij}$ can be expressed as,
 	\begin{align}
		A &= \Psi + \Psi^2 + \ddot{\xi} + {\cal H}\dot{\xi}, \\
		B_i &= \omega_i - \dot{\xi}_{\perp i}, \\
		D &= \Phi + \Phi^2 + \frac{\chi^{kl}\chi_{kl}}{3} - \frac{\Delta \xi}{3} + {\cal H}\dot{\xi}, \\
		E_{ij} &= (1+2\Phi)\chi_{ij} + \chi_i^k \chi_{kj} - \frac{\chi^{kl}\chi_{kl}}{3}\delta_{ij} - \left( \pd_i\pd_j - \frac{\Delta}{3}\delta_{ij} \right)\xi - \frac{\pd_i \xi_{\perp j} + \pd_j \xi_{\perp i}}{2}, 
	\end{align}
to second order, where ${\cal H}$ is the conformal Hubble parameter. 
Inversely, 
solving these equations, 
the metric perturbations $\Psi$, $\Phi$, $\omega^i$, and $\chi_{ij}$ can be written in terms of $A$, $B_i$, $D$, and $E_{ij}$ as,
	\begin{align}
		\Psi &= A - A^2 - \ddot{\xi} - {\cal H}\dot{\xi}, \\
		\omega_i &= B_i + \dot{\xi}_{\perp i}, \\
		\Phi &= D - D^2 - \frac{E^{kl}E_{kl} }{3} + \frac{\Delta \xi}{3} - {\cal H}\dot{\xi}, \\
		\chi_{ij} &= (1+2D)E_{ij} - E_i^k E_{kj} + \frac{E^{kl}E_{kl}}{3}\delta_{ij} + \left( \pd_i\pd_j - \frac{\Delta}{3}\delta_{ij} \right)\xi + \frac{\pd_i \xi_{\perp j} + \pd_j \xi_{\perp i}}{2} ,
	\end{align}
where, now, $\xi$ and $\xi_{\perp}^i$ are the solutions of Eqs. (\ref{eq:xi}) and (\ref{eq:xi_perp}) with 
	\begin{align}
		\pd^ i \hat{\gamma}_{ij} = 4E_{ij}\pd^i D + 2\pd^i\left( E_i^k E_{kj} - \frac{E^{kl}E_{kl}}{3}\delta_{ij} \right) .
	\end{align}
Note that the difference between two definitions are of second order. 
As for the difference from the definition in Ref. \cite{Huang:2013qua}, 
where the scalar modes are defined as ours, 
correction terms contain the first-order tensor modes.
Hence, 
their contributions are suppressed by the tensor-to-scalar ratio, $r < {\cal O}(0.1)$ \cite{Ade:2013uln, Ade:2014xna}, in comparison to the other second-order contributions. 

The momentum and the intensity are also changed by the gauge transformation as \cite{Naruko:2013aaa}, 
	\begin{align}
		q &\to q\left( 1 - {\cal H}\dot{\xi} - n^{(i)}_0 \pd_i \dot{\xi} \right) ,\\
		n^{(i)} &\to n^{(i)} - S_0^{(i)(j)}\dot{\xi}_{,j} + \xi_{\perp}^{[i,j]}n_{0(j)} ,
	\end{align}
and 
	\begin{align}
		I \to I + ({\cal H}\dot{\xi} + n^{(i)}_0 \pd_i \dot{\xi})\pd_{\ln q}\bar{I} .
	\end{align}

\section{Brightness, temperature, and spectral distortions}\label{sec:bts}

In this section, we show the relation between the brightness, temperature, and Compton $y$-parameter.
The distribution function is provided by the Planck distribution at zeroth order, 
	\begin{align}
		\bar{I} = I_{\rm BB}\left(\frac{q}{a\ol{T}}\right); \quad I_{\rm BB}(x) := \frac{2}{e^x -1} ,
	\end{align}
where background quantities are denoted with a bar. 
At second order, 
on the other hand, 
it deviates from the Planck distribution.
The deviation can be shown to be well approximated by the Compton $y$ parameter \cite{Pitrou:2009bc}, 
which is defined through the Fokker-Planck expansion as \cite{Stebbins:2007ve},
	\begin{align}\label{def:y}
		I \simeq I_{\rm BB}\left(\frac{q}{aT}\right) + y({\bs n}){\cal D}_q^2 I_{\rm BB}\left(\frac{q}{aT}\right); \quad {\cal D}_q \equiv q^{-3}\pdif{}{\ln q}q^3\pdif{}{\ln q}.
	\end{align}
Here, the temperature has also a deviation from that in the background,
	\begin{align}\label{def:temperature}
		T(\eta, {\bs n}) = \ol{T}(\eta)\left[1 + \Theta({\bs n}) \right].
	\end{align}
In this paper, 
instead of the temperature fluctuations $\Theta$, 
we have used the fractional brightness defined through,
	\begin{align}
		\Delta \equiv \frac{\delta B}{\bar{B}}; \quad B \equiv \int\!{\rm d}q~ q^3 I,
	\end{align}
to characterise the intensity $I$.
An advantage to use the fractional brightness is that it is linear in the intensity.
Hence, 
it is easy to derive its line-of-sight formula from that for the intensity.
In addition, 
its higher multipole moments do not appear in the energy-momentum tensor of the radiation 
while all multipole moments do in the case of the temperature fluctuations.

Using the definition in Eq. (\ref{def:y}), 
we can find a relation between the brightness, temperature, and Compton $y$-parameter as,
	\begin{align}\label{eq:bty}
		B = \frac{2\pi^4a^4}{15}T^4(1+4y) = \bar{B}(1+\Theta)^4(1+4y).
	\end{align}
Then, up to second order, we obtain
	\begin{align}
		\Delta = 4\Theta + 6\Theta^2 + 4y,
	\end{align}
and solving it perturbatively,
	\begin{align}
		\Theta^{(I)} &= \frac{1}{4}\Delta^{(I)},\\
		\Theta^{(II)} &= \frac{1}{4}\Delta^{(II)} - \frac{3}{32}\left(\Delta^{(I)}\right)^2 - 4y^{(II)},
	\end{align}
where we have used $y^{(I)}=0$. 

Finally, the energy-momentum tensor of the radiation can be written in terms of the brightness as,
	\begin{align}
		T^{(0)(0)} &= \frac{1}{a^4}\int\!{\rm d}^2n B , \\
		T^{(0)(i)} &= \frac{1}{a^4}\int\!{\rm d}^2n \left( n^{(i)} B \right),\\
		T^{(i)(j)} &= \frac{1}{a^4}\int\!{\rm d}^2n \left( n^{(i)}n^{(j)} B \right).
	\end{align}
As mentioned above, it can be written in terms of the multipole moments of the brightness up to quadrupole. 

\section{The line-of-sight formula by Huang \& Vernizzi}\label{sec:Huang}
Here, we review how the difficulty discussed in the subsection \ref{ss:difficulty} has been solved in Refs. \cite{Huang:2012ub, Huang:2013qua}. 
\footnote{We do not follow their derivation literally but only the flow of it.}
To show it, 
we go back to the original brightness equation,
	\begin{align}\label{eq:brightness_app}
		\dot{\Delta} + n^{(i)}\pd_i \Delta = \dot{\tau}C^{\Delta} + {\mathfrak D}^{\Delta} .
	\end{align}
First, they eliminated the $\Delta$-dependence in the redshift term by dividing the both sides by $1+\Delta$. 
In fact, 
introducing the new variable,
	\begin{align}
		\Delta_{\rm HV} \equiv \ln(1+\Delta) \simeq \Delta - \frac{\Delta^2}{2} ,
	\end{align}
the brightness equation can be rewritten as,
	\begin{align}\label{eq:brightness_HV}
		\dot{\Delta}_{\rm HV} + n^{(i)}\pd_i \Delta_{\rm HV} = \dot{\tau}C^{\Delta}_{\rm HV} + {\mathfrak D}^{\Delta}_{\rm HV} ,
	\end{align}
where
	\begin{align}
		C^{\Delta}_{\rm HV} \equiv \frac{C^{\Delta}}{1+\Delta} ,
	\end{align}
and
	\begin{align}
		{\mathfrak D}^{\Delta}_{\rm HV} \equiv 4D^q - \left( D^i \pd_i + D^{n^{(i)}}\pd_{n^{(i)}} \right)\Delta_{\rm HV} .
	\end{align}
Thus, 
the redshift term becomes independent of $\Delta$.

As Eq. (\ref{eq:LOSbg_x}), 
we rewrite Eq. (\ref{eq:brightness_HV}) in a line-of-sight integral form,
	\begin{align}
		\Delta_{\rm HV}(\eta_0, x_0^i,n_0^{(i)}) &= \int_{0}^{\eta_0}{\rm d}\eta' \left\{ e^{-\tau}\left[ 4D^q - \left( D^i \pd_i + D^{n^{(i)}}\pd_{n^{(i)}} \right)\Delta_{\rm HV} \right] + g_v S_{\rm HV}^{\Delta} \right\} ,
	\end{align}
where $S^{\Delta}_{\rm HV} \equiv \Delta_{\rm HV} - C_{\rm HV}^{\Delta}$.
Introducing the derivative $\dif{}{n^{(i)}} \equiv \pd_{n^{(i)}} + (\eta-\eta_0)\pd_i$, which commutes with $\pd_{\eta} + n^{(i)}\pd_i$, 
the time-delay and lensing terms becomes,
	\begin{align}
		\left( D^i \pd_i + D^{n^{(i)}}\pd_{n^{(i)}} \right)\Delta_{\rm HV} =
		 \left\{ \left[ D^i - (\eta-\eta_0)D^{n^{(i)}} \right]\pd_i + D^{n^{(i)}}\dif{}{n^{(i)}} \right\}\Delta_{\rm HV} .
	\end{align}
Then, using the primitive functions,
	\begin{align}
		X^i(\eta') &\equiv \int_0^{\eta'}{\rm d}\eta_1  e^{-\tau} \left[ D^i(\eta_1, \bar{x}^i, \bar{n}^{(i)}) - (\eta_1-\eta_0)D^{n^{{i}}}(\eta_1, \bar{x}^i, \bar{n}^{(i)}) \right], \\
		N^{(i)}(\eta') &\equiv \int_0^{\eta'}{\rm d}\eta_1  e^{-\tau} D^{n^{{i}}}(\eta_1, \bar{x}^i, \bar{n}^{(i)}) ,
	\end{align}
we can perform an integration by parts as,
	\begin{align}
		\int_{0}^{\eta_0}{\rm d}\eta' e^{-\tau}\left[ \left( D^i \pd_i + D^{n^{(i)}}\pd_{n^{(i)}} \right)\Delta_{\rm HV} \right]  
		&= \left[ X^i(\eta_0) \pd_i + N^{(i)}(\eta_0)\pd_{n^{(i)}} \right]\Delta_{\rm HV}(\eta_0) \nm \\
		& - \int_{0}^{\eta_0}{\rm d}\eta' e^{-\tau}\left[ \left( X^i \pd_i + N^{(i)}\dif{}{n^{(i)}} \right)(\pd_\eta + n_0^{(i)}\pd_i)\Delta_{\rm HV} \right] .
	\end{align}
Using the modified brightness equation (\ref{eq:brightness_HV}), 
the second term can be rewritten as
	\begin{align}
		&\int_{0}^{\eta_0}{\rm d}\eta' e^{-\tau}\left[ \left( X^i \pd_i + N^{(i)}\dif{}{n^{(i)}} \right)(\pd_\eta + n_0^{(i)}\pd_i)\Delta_{\rm HV} \right] = \nm \\
		& \hspace*{.3\linewidth} \int_{0}^{\eta_0}{\rm d}\eta' e^{-\tau}\left[ \left( X^i \pd_i + N^{(i)}\dif{}{n^{(i)}} \right)(\dot{\tau}C^{\Delta}_{\rm HV} + {\mathfrak D}^{\Delta}_{\rm HV}) \right] .
	\end{align}
Again, the integrand depends on $\Delta$ due to the time-delay and lensing terms in ${\mathfrak D}^{\Delta}_{\rm HV}$ 
but their contributions are of higher order in the perturbations.
The remaining terms only depends on a few low multipole moments of $\Delta_{\rm HV}$.

\section{Generalization of the SW-ISW decomposition to higher orders}\label{sec:GE}
Here, we derive the higher-order geodesic equations and show how the decomposition into the Sachs-Wolfe (SW) and the integrated Sachs-Wolfe (ISW) terms can be generalized to higher orders. 

The full geodesic equations are given by,
	\begin{align}
		\dif{x^i}{\lambda} &= p^i = \frac{\itetrad{a}{i}q^{(a)}}{a}, \label{eq:fullgeodesic_x} \\
		\dif{q_{(a)}}{\lambda} &= \frac{2\spin{a}{b}{c}q_{(b)}q^{(c)}}{a^2}; \quad \spin{a}{b}{c} \equiv \ctetrad{b}{[\mu,\nu]}\citetrad{c}{\mu}\citetrad{a}{\nu}, \label{eq:fullgeodesic_q}
	\end{align}
where $\lambda$ is an affine parameter of the geodesic curve, which is related to the conformal time as,
	\begin{align}
		\dif{\eta}{\lambda} = p^0 = \frac{\itetrad{a}{0}q^{(a)}}{a} . \label{eq:fullgeodesic_eta}
	\end{align}
Here, 
we have introduced the tetrad and its inverse for the rescaled metric $\tilde{g}_{\mu\nu} \equiv g_{\mu\nu}/a^2$ by $\ctetrad{a}{\mu} \equiv \tetrad{a}{\mu}/a$ and $\citetrad{a}{\mu} \equiv a\itetrad{a}{\mu}$, respectively.
The latter equation (\ref{eq:fullgeodesic_q}) can be obtained from the geodesic equation $p_{\mu;\nu}p^{\nu}=0$ and the identity $p_{\nu;\mu}p^{\nu}=0$.

First, 
we derive the geodesic equations for the spacetime coordinates. 
Using the explicit form of the tetrad (\ref{eq:tetrad}) and its inverse (\ref{eq:itetrad}), 
the equations for the spacetime coordinates can be written as,
	\begin{align}
		\dif{\eta}{\lambda} &= \frac{qe^{-\Psi} }{a^2} , \\
		\dif{x^i}{\lambda} &= \frac{q}{a^2}\left([e^{-{\boldsymbol h}}]^i_j n^{(j)} - e^{-\Psi}\omega^i \right) ,
	\end{align}
and then
	\begin{align}
		\dif{x^i}{\eta} = e^{\Psi} \tilde{n}^i - \omega^i ,
	\end{align}
where
	\begin{align}
		\tilde{n}^\mu \equiv \citetrad{i}{\mu}n^{(i)} = [e^{-{\boldsymbol h}}]^j_i \delta^{\mu}_j n^{(i)}.
	\end{align}
In the first-order accuracy, 
we obtain the equation (\ref{eq:geodesic2nd_x}).

Next, since $q=q^{(0)}=-q_{(0)}$, 
the equation for $q$ is obtained from the 0-component of Eq. (\ref{eq:fullgeodesic_q}) as
	\begin{align}
		\dif{q}{\lambda} &= -\frac{2\spinl{0}{b}{c}q^{(b)}q^{(c)}}{a^2} = -\frac{2q^2}{a^2}(\spinl{0}{0}{i}n^{(i)} + \spinl{0}{i}{j}n^{(i)}n^{(j)}) .
	\end{align}
First, 
$\spinl{0}{0}{i}$ can be written as,
	\begin{align}
		\spinl{0}{0}{i} = \frac{\Psi_{,j}}{2}[e^{-{\boldsymbol h}}]^j_i.
	\end{align}
As for $\spinl{0}{i}{j}$, only its symmetric part is necessary. 
Using the identity $\tetrad{b}{[\mu,\nu]}=\tetrad{b}{[\mu;\nu]}$ and $\tetrad{b}{\mu;\nu}\itetrad{c}{\mu}=-\tetrad{b}{\mu}{\itetrad{c}{\mu}}_{;\nu}$, 
we find
	\begin{align}
		\frac{\spinl{0}{i}{j} + \spinl{0}{j}{i}}{2} = \frac{1}{2}\tilde{e}_{(0)(\mu;\nu)}\citetrad{i}{\mu}\citetrad{j}{\nu} .
	\end{align}
Since we have chosen the tetrad so that the vector $\itetrad{0}{\mu}$ is normal to the constant time hypersurfaces, 
this can be written in terms of the extrinsic curvature for the rescaled metric $\tilde{K}_{ij}$ as,
	\begin{align}
		\frac{\spinl{0}{i}{j} + \spinl{0}{j}{i}}{2} = \frac{1}{2}\tilde{K}_{kl}[e^{-{\boldsymbol h}}]^k_i [e^{-{\boldsymbol h}}]^l_j \,,
	\end{align}
From these equalities, 
we get
	\begin{align}
		\dif{\ln q}{\eta} = -e^{\Psi}(\Psi_{,j} \tilde{n}^i + \tilde{K}_{ij} \tilde{n}^i \tilde{n}^j) .
	\end{align}
	
As in the second-order case, 
the first term in the RHS can be written in terms of ${\rm d}x^i/{\rm d}\eta$. 
Then, 
we get the generalization of the decomposition into the SW and ISW terms:
	\begin{align}
		\dif{}{\eta}\left(\ln q + \Psi \right) &=  \dot{\Psi} - \omega^i \Psi_{,i} - e^{\Psi}\tilde{K}_{ij} \tilde{n}^i \tilde{n}^j . \label{eq:fullSWISW}
	\end{align}
The right-hand side of this equation can be further rewritten in terms of the Lie derivative along the normal vector. 
The expressions become concise when one uses the {\it non-unit} normal vector, 
	\begin{align}\label{eq:normal_N}
		N^{\mu} \equiv e^{\Psi}\citetrad{0}{\mu} = \delta^\mu_0-\omega^i\delta^\mu_i .
	\end{align}
Using the Lie derivative along the vector $N^{\mu}$, 
it is easy to see that the first term can be rewritten as
	\begin{align}
		\dot{\Psi} - \omega^i \Psi_{,i} = {\cal L}_N (\Psi).
	\end{align}
The second term can be also written as
	\begin{align}
		e^{\Psi}\tilde{K}_{ij} \tilde{n}^i \tilde{n}^j &= \frac{e^{\Psi}}{2}{\cal L}_{\tilde{e}_{(0)}}(\tilde{g}_{\mu\nu})\tilde{n}^{\mu} \tilde{n}^{\nu} \\
		&= \frac{e^{\Psi}}{2}\left[ {\cal L}_{\tilde{e}_{(0)}}(\tilde{e}_{(i)\mu})\citetrad{j}{\mu} + (i \leftrightarrow j) \right] n^{(i)}n^{(j)} ,
	\end{align}
using the Lie derivative along the normal vector $\citetrad{0}{\mu}$. 
Substituting the explicit form (\ref{eq:tetrad}) and (\ref{eq:itetrad}), 
it can be further rewritten as,
	\begin{align}
		e^{\Psi}{\cal L}_{\tilde{e}_{(0)}}(\tilde{e}_{(i)\mu})\citetrad{j}{\mu} &= [e^{\bs h}]_{ik,\mu}N^{\mu}[e^{-{\bs h}}]^k_j + [e^{\bs h}]_{ik}N^k_{,l}[e^{-{\bs h}}]^l_j \\
		&= \delta_{ik} N^k_{,j} + \left[\sum_{n=0}^{\infty}\frac{{\rm ad}({\bs h})^n}{(n+1)!}{\bs k}({\bs h}) \right]_{ij}
	\end{align}
where the operator ${\bs k}$ is defined for a tensor on a constant time hypersurface as
	\begin{align}
		k_{ij}({\bs h}) &\equiv {\cal L}_N(h^{\mu}_{\nu})\delta_{\mu i}\delta^\nu_j; \quad h^{\mu}_{\nu} \equiv h^i_j \delta_i^{\mu} \delta^j_{\nu} \\
		&= h_{ij,\mu}N^{\mu} + h_{ik}N^k_{,j} - N_{i,k}h^k_j,
	\end{align}
and ${\rm ad}({\bs h})$ is an operator that maps a matrix to a commutator as
	\begin{align}
		{\rm ad}({\bs h}): * \to [{\bs h}, *].
	\end{align}
Decomposing the quantities into the scalar, vector, and tensor modes, 
we obtain
	\begin{align}
		e^{\Psi}{\cal L}_{\tilde{e}_{(0)}}(\tilde{e}_{(i)\mu})\citetrad{j}{\mu} &= {\cal L}_N(\Phi) + \omega_{i,j} + \left[\sum_{n=0}^{\infty}\frac{{\rm ad}({\bs \chi})^n}{(n+1)!}{\bs k}({\bs \chi}) \right]_{ij} .
	\end{align}
Substituting the results into Eq. (\ref{eq:fullSWISW}), 
we get explicit form of the generalized decomposition into the SW and ISW contributions:
	\begin{align}
		\dif{}{\eta}\left(\ln q + \Psi \right) &= {\cal L}_N(\Psi-\Phi) - \left( \omega_{(i,j)} + \left[\sum_{n=0}^{\infty}\frac{{\rm ad}({\bs \chi})^{2n}}{(2n+1)!}{\bs k}({\bs \chi}) \right]_{ij} \right) n^{(i)} n^{(j)} . \label{eq:fullSWISW_SVT}
	\end{align}
In the second-order accuracy, 
this reproduces the second-order result (\ref{eq:SWISW}).
It is remarkable that the terms of the scalar modes have similar to those in the linear case 
except that a time derivative is replaced by the Lie derivative along the normal vector $N^{\mu}$. 
This is not the case if the standard definition (\ref{def:metric_std}) is employed for the vector and tensor modes. 
The corresponding formula for the standard definition can be obtained by making the replacements,
	\begin{align}
		\Psi &\to \frac{1}{2}\ln(1+2A + \gamma^{ij}B_iB_j) = A - A^2 + \cdots, \\ 
		\Phi &\to \frac{1}{2}\ln(1+2D) = D - D^2 + \cdots, \\
		\omega^i &\to \gamma^{ij}B_j = (1-2D)B^i - 2E^{ij}B_j + \cdots, \\
		\chi_{ij} &\to \frac{1}{2}\left[\ln\left({\bs I} + \frac{2{\bs E}}{1+2D} \right)\right]_{ij} = (1-2D)E_{ij} - E_{ik}E^k_j + \cdots ,
	\end{align}
because we have not used the gauge conditions $\omega^i_{,i}=0$ and $\chi^i_{j,i}=0$ in the derivation.

Finally, 
we close this Appendix by giving the geodesic equation for the direction $n^{(i)}$:
	\begin{align}
		\dif{n^{(i)}}{\eta} &= 2e^{\Psi}S^{(i)(j)}\left[ \omega_{(j)(0)(0)} + (\omega_{(j)(k)(0)} + \omega_{(j)(0)(k)})n^{(k)} + \omega_{(j)(k)(l)}n^{(k)}n^{(l)} \right] \\
		&= -S^{(i)(j)}\tilde{\pd}_{(j)}(\Psi-\Phi) - S^{(i)(j)}n^{(k)}\left[ \sum_{n=0}^{\infty}\frac{{\rm ad}({\bs \chi})^{n}}{(n+1)!} \tilde{\pd}_{(0)}{\bs \chi} \right]_{kj} \nm \\
		&\hspace*{.3\linewidth} - 2S^{(i)[(j)}n^{(l)]}n^{(k)}\left[ \sum_{n=0}^{\infty}\frac{{\rm ad}({\bs \chi})^{n}}{(n+1)!} \tilde{\pd}_{(l)}{\bs \chi} \right]_{kj} ,
	\end{align}
where $\tilde{\pd}_{(a)}$ represents the directional derivative along $\citetrad{a}{\mu}$,
	\begin{align}
		\tilde{\pd}_{(0)} &\equiv \citetrad{0}{\mu}\pd_{\mu} = e^{-\Psi}(\pd_0 - \omega^i \pd_i) , \\
		\tilde{\pd}_{(i)} &\equiv \citetrad{i}{\mu}\pd_{\mu} = e^{-\Phi}[e^{-{\bs \chi}}]^j_i \pd_j .
	\end{align}
In the first-order accuracy, 
we obtain the equation (\ref{eq:geodesic2nd_n}).

\section{Line-of-sight formulae for the coupling terms}\label{sec:LOSc}
Here, we present the line-of-sight formulae for the coupling terms that have not shown in the subsection \ref{ss:LOS2nd}. These terms appear from the third term in Eq. (\ref{eq:mapping2nd}) and can be separated into the following five contributions:\\

- {\it SW $\times$ lensing} 

	\begin{align}
		\delta I_{{\rm SW} \times {\rm L}} &= \frac{\pd_{\ln q} \bar{I}(q)}{(2\pi)^6}\int {\rm d}^3k_1 \int_{0}^{\eta_0}\!{\rm d}\eta' \bar{g}_v(\eta') k_1^iT^{{\rm SW}\pI}({\bs k}_1, \eta') \nm \\
		&\qquad \qquad \times \left[ \int{\rm d}^3k_2 \int_{\eta'}^{\eta_0}\!{\rm d}\eta_1 T^{{\rm L}\pI}({\bs k}_2, \eta_1)e^{-i[{\bs k}_1(\eta_0-\eta')+{\bs k}_2(\eta_0-\eta_1)] \cdot {\bs n}_{\rm o}}  \right] . \label{eq:ISWL}
	\end{align}
	
- {\it SW $\times$ time delay} 

	\begin{align}
		\delta I_{{\rm SW} \times {\rm TD}} &= \frac{\pd_{\ln q} \bar{I}(q)}{(2\pi)^6}\int{\rm d}^3k_1 \int_{0}^{\eta_0}\!{\rm d}\eta' \bar{g}_v(\eta') k_1^iT^{{\rm SW}\pI}({\bs k}_1, \eta') \nm \\
		&\qquad \qquad \times \left[ \int{\rm d}^3k_2 \int_{\eta'}^{\eta_0}\!{\rm d}\eta_1 T^{{\rm TD}\pI}({\bs k}_2, \eta_1)e^{-i[{\bs k}_1(\eta_0-\eta')+{\bs k}_2(\eta_0-\eta_1)] \cdot {\bs n}_{\rm o}}  \right] . \label{eq:ISWTD}
	\end{align}
	
- {\it ISW $\times$ lensing} 

	\begin{align}
		\delta I_{{\rm ISW} \times {\rm L}} &= \frac{\pd_{\ln q} \bar{I}(q)}{(2\pi)^6} \int{\rm d}^3k_1 \int_{0}^{\eta_0}\!{\rm d}\eta_2 \tilde{T}^{{\rm L}\pI}({\bs k}_1, \eta_2) \nm \\
		&\qquad \times \left[ \int{\rm d}^3k_2 \int_{0}^{\eta_2}\!{\rm d}\eta_1 e^{-\bar{\tau}(\eta_1)}k_2^i(\eta_2-\eta_1)T^{{\rm ISW}\pI}({\bs k}_2, \eta_1)e^{-i[{\bs k}_1(\eta_0-\eta_2)+{\bs k}_2(\eta_0-\eta_1)] \cdot {\bs n}_{\rm o}}  \right] . \label{eq:IISWL}
	\end{align}

- {\it ISW $\times$ time delay} 

	\begin{align}
		\delta I_{{\rm ISW} \times {\rm TD}} &= \frac{\pd_{\ln q} \bar{I}(q)}{(2\pi)^6} \int{\rm d}^3k_1 \int_{0}^{\eta_0}\!{\rm d}\eta_2 T^{{\rm TD}\pI}({\bs k}_1, \eta_2) \nm \\
		&\qquad \qquad \times \left[ \int{\rm d}^3k_2 \int_{0}^{\eta_2}\!{\rm d}\eta_1 e^{-\bar{\tau}(\eta_1)}k_2^iT^{{\rm ISW}\pI}({\bs k}_2, \eta_1)e^{-i[{\bs k}_1(\eta_0-\eta_2)+{\bs k}_2(\eta_0-\eta_1)] \cdot {\bs n}_{\rm o}}  \right] . \label{eq:IISWISW}
	\end{align}

- {\it ISW $\times$ deflection} 
	
	\begin{align}
		\delta I_{{\rm ISW} \times {\rm D}} &= \frac{\pd_{\ln q} \bar{I}(q)}{(2\pi)^6} \int{\rm d}^3k_1 \int_{0}^{\eta_0}\!{\rm d}\eta_2 T^{{\rm D}\pI}({\bs k}_1, \eta_2) \nm \\
		&\qquad \qquad \times \left[ \int{\rm d}^3k_2 \int_{0}^{\eta_2}\!{\rm d}\eta_1 e^{-\bar{\tau}(\eta_1)}\pd_{n^{(i)}}T^{{\rm ISW}\pI}({\bs k}_2, \eta_1)e^{-i[{\bs k}_1(\eta_0-\eta_2)+{\bs k}_2(\eta_0-\eta_1)] \cdot {\bs n}_{\rm o}}  \right] . \label{eq:IISWISW}
	\end{align}


\providecommand{\href}[2]{#2}\begingroup\raggedright

\endgroup


\begin{thebibliography}{10}

\bibitem{Hinshaw:2012aka}
{\bf WMAP} Collaboration, G.~Hinshaw et~al., {\it {Nine-Year Wilkinson
  Microwave Anisotropy Probe (WMAP) Observations: Cosmological Parameter
  Results}},  {\em Astrophys.J.Suppl.} {\bf 208} (2013) 19,
  [\href{http://xxx.lanl.gov/abs/1212.5226}{{\tt arXiv:1212.5226}}].

\bibitem{Hou:2012xq}
Z.~Hou, C.~Reichardt, K.~Story, B.~Follin, R.~Keisler, et~al., {\it
  {Constraints on Cosmology from the Cosmic Microwave Background Power Spectrum
  of the 2500-square degree SPT-SZ Survey}},  {\em Astrophys.J.} {\bf 782}
  (2014) 74, [\href{http://xxx.lanl.gov/abs/1212.6267}{{\tt arXiv:1212.6267}}].

\bibitem{Sievers:2013ica}
{\bf Atacama Cosmology Telescope} Collaboration, J.~L. Sievers et~al., {\it
  {The Atacama Cosmology Telescope: Cosmological parameters from three seasons
  of data}},  {\em JCAP} {\bf 1310} (2013) 060,
  [\href{http://xxx.lanl.gov/abs/1301.0824}{{\tt arXiv:1301.0824}}].

\bibitem{Calabrese:2013jyk}
E.~Calabrese, R.~A. Hlozek, N.~Battaglia, E.~S. Battistelli, J.~R. Bond,
  et~al., {\it {Cosmological parameters from pre-planck cosmic microwave
  background measurements}},  {\em Phys.Rev.} {\bf D87} (2013), no.~10 103012,
  [\href{http://xxx.lanl.gov/abs/1302.1841}{{\tt arXiv:1302.1841}}].

\bibitem{Ade:2013zuv}
{\bf Planck} Collaboration, P.~Ade et~al., {\it {Planck 2013 results. XVI.
  Cosmological parameters}},  \href{http://xxx.lanl.gov/abs/1303.5076}{{\tt
  arXiv:1303.5076}}.

\bibitem{Ade:2013uln}
{\bf Planck} Collaboration, P.~Ade et~al., {\it {Planck 2013 results. XXII.
  Constraints on inflation}},  \href{http://xxx.lanl.gov/abs/1303.5082}{{\tt
  arXiv:1303.5082}}.

\bibitem{Kodama:1985bj}
H.~Kodama and M.~Sasaki, {\it {Cosmological Perturbation Theory}},  {\em
  Prog.Theor.Phys.Suppl.} {\bf 78} (1984) 1--166.

\bibitem{Mukhanov:1990me}
V.~F. Mukhanov, H.~Feldman, and R.~H. Brandenberger, {\it {Theory of
  cosmological perturbations. Part 1. Classical perturbations. Part 2. Quantum
  theory of perturbations. Part 3. Extensions}},  {\em Phys.Rept.} {\bf 215}
  (1992) 203--333.

\bibitem{Lewis:2006fu}
A.~Lewis and A.~Challinor, {\it {Weak gravitational lensing of the cmb}},  {\em
  Phys.Rept.} {\bf 429} (2006) 1--65,
  [\href{http://xxx.lanl.gov/abs/astro-ph/0601594}{{\tt astro-ph/0601594}}].

\bibitem{Hanson:2013hsb}
{\bf SPTpol} Collaboration, D.~Hanson et~al., {\it {Detection of B-mode
  Polarization in the Cosmic Microwave Background with Data from the South Pole
  Telescope}},  {\em Phys.Rev.Lett.} {\bf 111} (2013), no.~14 141301,
  [\href{http://xxx.lanl.gov/abs/1307.5830}{{\tt arXiv:1307.5830}}].

\bibitem{Ade:2013hjl}
{\bf POLARBEAR} Collaboration, P.~Ade et~al., {\it {Evidence for Gravitational
  Lensing of the Cosmic Microwave Background Polarization from
  Cross-correlation with the Cosmic Infrared Background}},  {\em
  Phys.Rev.Lett.} {\bf 112} (2014) 131302,
  [\href{http://xxx.lanl.gov/abs/1312.6645}{{\tt arXiv:1312.6645}}].

\bibitem{Ade:2013gez}
{\bf POLARBEAR} Collaboration, P.~Ade et~al., {\it {Measurement of the Cosmic
  Microwave Background Polarization Lensing Power Spectrum with the POLARBEAR
  experiment}},  {\em Phys.Rev.Lett.} {\bf 113} (2014) 021301,
  [\href{http://xxx.lanl.gov/abs/1312.6646}{{\tt arXiv:1312.6646}}].

\bibitem{Ade:2014afa}
{\bf The POLARBEAR} Collaboration, P.~Ade et~al., {\it {A Measurement of the
  Cosmic Microwave Background B-Mode Polarization Power Spectrum at Sub-Degree
  Scales with POLARBEAR}},  \href{http://xxx.lanl.gov/abs/1403.2369}{{\tt
  arXiv:1403.2369}}.

\bibitem{Bartolo:2006cu}
N.~Bartolo, S.~Matarrese, and A.~Riotto, {\it {CMB Anisotropies at Second Order
  I}},  {\em JCAP} {\bf 0606} (2006) 024,
  [\href{http://xxx.lanl.gov/abs/astro-ph/0604416}{{\tt astro-ph/0604416}}].

\bibitem{Bartolo:2006fj}
N.~Bartolo, S.~Matarrese, and A.~Riotto, {\it {CMB Anisotropies at
  Second-Order. 2. Analytical Approach}},  {\em JCAP} {\bf 0701} (2007) 019,
  [\href{http://xxx.lanl.gov/abs/astro-ph/0610110}{{\tt astro-ph/0610110}}].

\bibitem{Pitrou:2008hy}
C.~Pitrou, {\it {The Radiative transfer at second order: A Full treatment of
  the Boltzmann equation with polarization}},  {\em Class.Quant.Grav.} {\bf 26}
  (2009) 065006, [\href{http://xxx.lanl.gov/abs/0809.3036}{{\tt
  arXiv:0809.3036}}].

\bibitem{Pitrou:2008ut}
C.~Pitrou, {\it {The radiative transfer for polarized radiation at second order
  in cosmological perturbations}},  {\em Gen.Rel.Grav.} {\bf 41} (2009)
  2587--2595, [\href{http://xxx.lanl.gov/abs/0809.3245}{{\tt
  arXiv:0809.3245}}].

\bibitem{Beneke:2010eg}
M.~Beneke and C.~Fidler, {\it {Boltzmann hierarchy for the cosmic microwave
  background at second order including photon polarization}},  {\em Phys.Rev.}
  {\bf D82} (2010) 063509, [\href{http://xxx.lanl.gov/abs/1003.1834}{{\tt
  arXiv:1003.1834}}].

\bibitem{Pitrou:2007jy}
C.~Pitrou, {\it {Gauge invariant Boltzmann equation and the fluid limit}},
  {\em Class.Quant.Grav.} {\bf 24} (2007) 6127--6158,
  [\href{http://xxx.lanl.gov/abs/0706.4383}{{\tt arXiv:0706.4383}}].

\bibitem{Naruko:2013aaa}
A.~Naruko, C.~Pitrou, K.~Koyama, and M.~Sasaki, {\it {Second-order Boltzmann
  equation: gauge dependence and gauge invariance}},  {\em Class.Quant.Grav.}
  {\bf 30} (2013) 165008, [\href{http://xxx.lanl.gov/abs/1304.6929}{{\tt
  arXiv:1304.6929}}].

\bibitem{Khatri:2008kb}
R.~Khatri and B.~D. Wandelt, {\it {Crinkles in the last scattering surface:
  Non-Gaussianity from inhomogeneous recombination}},  {\em Phys.Rev.} {\bf
  D79} (2009) 023501, [\href{http://xxx.lanl.gov/abs/0810.4370}{{\tt
  arXiv:0810.4370}}].

\bibitem{Bartolo:2008sg}
N.~Bartolo and A.~Riotto, {\it {On the non-Gaussianity from Recombination}},
  {\em JCAP} {\bf 0903} (2009) 017,
  [\href{http://xxx.lanl.gov/abs/0811.4584}{{\tt arXiv:0811.4584}}].

\bibitem{Senatore:2008vi}
L.~Senatore, S.~Tassev, and M.~Zaldarriaga, {\it {Cosmological Perturbations at
  Second Order and Recombination Perturbed}},  {\em JCAP} {\bf 0908} (2009)
  031, [\href{http://xxx.lanl.gov/abs/0812.3652}{{\tt arXiv:0812.3652}}].

\bibitem{Senatore:2008wk}
L.~Senatore, S.~Tassev, and M.~Zaldarriaga, {\it {Non-Gaussianities from
  Perturbing Recombination}},  {\em JCAP} {\bf 0909} (2009) 038,
  [\href{http://xxx.lanl.gov/abs/0812.3658}{{\tt arXiv:0812.3658}}].

\bibitem{Nitta:2009jp}
D.~Nitta, E.~Komatsu, N.~Bartolo, S.~Matarrese, and A.~Riotto, {\it {CMB
  anisotropies at second order III: bispectrum from products of the first-order
  perturbations}},  {\em JCAP} {\bf 0905} (2009) 014,
  [\href{http://xxx.lanl.gov/abs/0903.0894}{{\tt arXiv:0903.0894}}].

\bibitem{Boubekeur:2009uk}
L.~Boubekeur, P.~Creminelli, G.~D'Amico, J.~Norena, and F.~Vernizzi, {\it
  {Sachs-Wolfe at second order: the CMB bispectrum on large angular scales}},
  {\em JCAP} {\bf 0908} (2009) 029,
  [\href{http://xxx.lanl.gov/abs/0906.0980}{{\tt arXiv:0906.0980}}].

\bibitem{Gao:2010ti}
X.~Gao, {\it {On non-linear CMB temperature anisotropy from gravitational
  perturbations}},  {\em Phys.Rev.} {\bf D82} (2010) 103004,
  [\href{http://xxx.lanl.gov/abs/1005.1219}{{\tt arXiv:1005.1219}}].

\bibitem{Creminelli:2011sq}
P.~Creminelli, C.~Pitrou, and F.~Vernizzi, {\it {The CMB bispectrum in the
  squeezed limit}},  {\em JCAP} {\bf 1111} (2011) 025,
  [\href{http://xxx.lanl.gov/abs/1109.1822}{{\tt arXiv:1109.1822}}].

\bibitem{Bartolo:2011wb}
N.~Bartolo, S.~Matarrese, and A.~Riotto, {\it {Non-Gaussianity in the Cosmic
  Microwave Background Anisotropies at Recombination in the Squeezed limit}},
  {\em JCAP} {\bf 1202} (2012) 017,
  [\href{http://xxx.lanl.gov/abs/1109.2043}{{\tt arXiv:1109.2043}}].

\bibitem{Pitrou:2010sn}
C.~Pitrou, J.-P. Uzan, and F.~Bernardeau, {\it {The cosmic microwave background
  bispectrum from the non-linear evolution of the cosmological perturbations}},
   {\em JCAP} {\bf 1007} (2010) 003,
  [\href{http://xxx.lanl.gov/abs/1003.0481}{{\tt arXiv:1003.0481}}].

\bibitem{Su:2012gt}
S.-C. Su, E.~A. Lim, and E.~Shellard, {\it {CMB Bispectrum from Non-linear
  Effects during Recombination}},
  \href{http://xxx.lanl.gov/abs/1212.6968}{{\tt arXiv:1212.6968}}.

\bibitem{Huang:2012ub}
Z.~Huang and F.~Vernizzi, {\it {Cosmic Microwave Background Bispectrum from
  Recombination}},  {\em Phys.Rev.Lett.} {\bf 110} (2013), no.~10 101303,
  [\href{http://xxx.lanl.gov/abs/1212.3573}{{\tt arXiv:1212.3573}}].

\bibitem{Pettinari:2013he}
G.~W. Pettinari, C.~Fidler, R.~Crittenden, K.~Koyama, and D.~Wands, {\it {The
  intrinsic bispectrum of the Cosmic Microwave Background}},  {\em JCAP} {\bf
  1304} (2013) 003, [\href{http://xxx.lanl.gov/abs/1302.0832}{{\tt
  arXiv:1302.0832}}].

\bibitem{Huang:2013qua}
Z.~Huang and F.~Vernizzi, {\it {The full CMB temperature bispectrum from
  single-field inflation}},  {\em Phys.Rev.} {\bf D89} (2014) 021302,
  [\href{http://xxx.lanl.gov/abs/1311.6105}{{\tt arXiv:1311.6105}}].

\bibitem{Seljak:1996is}
U.~Seljak and M.~Zaldarriaga, {\it {A Line of sight integration approach to
  cosmic microwave background anisotropies}},  {\em Astrophys.J.} {\bf 469}
  (1996) 437--444, [\href{http://xxx.lanl.gov/abs/astro-ph/9603033}{{\tt
  astro-ph/9603033}}].

\bibitem{Fidler:2014oda}
C.~Fidler, G.~W. Pettinari, M.~Beneke, R.~Crittenden, K.~Koyama, et~al., {\it
  {The intrinsic B-mode polarisation of the Cosmic Microwave Background}},
  \href{http://xxx.lanl.gov/abs/1401.3296}{{\tt arXiv:1401.3296}}.

\bibitem{Pettinari:2014iha}
G.~W. Pettinari, C.~Fidler, R.~Crittenden, K.~Koyama, A.~Lewis, et~al., {\it
  {Impact of polarisation on the intrinsic CMB bispectrum}},
  \href{http://xxx.lanl.gov/abs/1406.2981}{{\tt arXiv:1406.2981}}.

\bibitem{Su:2014mga}
S.~C. Su and E.~A. Lim, {\it {Formulating Weak Lensing from the Boltzmann
  Equation and Application to Lens-lens Couplings}},  {\em Phys.Rev.} {\bf D89}
  (2014) 123006, [\href{http://xxx.lanl.gov/abs/1401.5737}{{\tt
  arXiv:1401.5737}}].

\bibitem{Misner:1974qy}
C.~W. Misner, K.~Thorne, and J.~Wheeler, {\em {Gravitation}}.
\newblock W.H.~Freeman and Co., 1974.

\bibitem{Arnowitt:1962hi}
R.~L. Arnowitt, S.~Deser, and C.~W. Misner, {\it {The Dynamics of general
  relativity}},  {\em Gen.Rel.Grav.} {\bf 40} (2008) 1997--2027,
  [\href{http://xxx.lanl.gov/abs/gr-qc/0405109}{{\tt gr-qc/0405109}}].

\bibitem{Maldacena:2002vr}
J.~M. Maldacena, {\it {Non-Gaussian features of primordial fluctuations in
  single field inflationary models}},  {\em JHEP} {\bf 0305} (2003) 013,
  [\href{http://xxx.lanl.gov/abs/astro-ph/0210603}{{\tt astro-ph/0210603}}].

\bibitem{Gao:2011vs}
X.~Gao, T.~Kobayashi, M.~Yamaguchi, and J.~Yokoyama, {\it {Primordial
  non-Gaussianities of gravitational waves in the most general single-field
  inflation model}},  {\em Phys.Rev.Lett.} {\bf 107} (2011) 211301,
  [\href{http://xxx.lanl.gov/abs/1108.3513}{{\tt arXiv:1108.3513}}].

\bibitem{Gao:2012ib}
X.~Gao, T.~Kobayashi, M.~Shiraishi, M.~Yamaguchi, J.~Yokoyama, et~al., {\it
  {Full bispectra from primordial scalar and tensor perturbations in the most
  general single-field inflation model}},  {\em PTEP} {\bf 2013} (2013) 053E03,
  [\href{http://xxx.lanl.gov/abs/1207.0588}{{\tt arXiv:1207.0588}}].

\bibitem{Challinor:2000as}
A.~Challinor, {\it {Microwave background polarization in cosmological models}},
   {\em Phys.Rev.} {\bf D62} (2000) 043004,
  [\href{http://xxx.lanl.gov/abs/astro-ph/9911481}{{\tt astro-ph/9911481}}].

\bibitem{Komatsu:2001rj}
E.~Komatsu and D.~N. Spergel, {\it {Acoustic signatures in the primary
  microwave background bispectrum}},  {\em Phys.Rev.} {\bf D63} (2001) 063002,
  [\href{http://xxx.lanl.gov/abs/astro-ph/0005036}{{\tt astro-ph/0005036}}].

\bibitem{1953ApJ...117..134L}
D.~N. {Limber}, {\it {The Analysis of Counts of the Extragalactic Nebulae in
  Terms of a Fluctuating Density Field.}},  {\em Astrophys.J.} {\bf 117} (1953)
  134.

\bibitem{Fidler:2014zwa}
C.~Fidler, K.~Koyama, and G.~W. Pettinari, {\it {A new line-of-sight approach
  to the non-linear Cosmic Microwave Background}},
  \href{http://xxx.lanl.gov/abs/1409.2461}{{\tt arXiv:1409.2461}}.

\bibitem{Ade:2014xna}
{\bf BICEP2} Collaboration, P.~Ade et~al., {\it {Detection of B-Mode
  Polarization at Degree Angular Scales by BICEP2}},  {\em Phys.Rev.Lett.} {\bf
  112} (2014) 241101, [\href{http://xxx.lanl.gov/abs/1403.3985}{{\tt
  arXiv:1403.3985}}].

\bibitem{Pitrou:2009bc}
C.~Pitrou, F.~Bernardeau, and J.-P. Uzan, {\it {The y-sky: diffuse spectral
  distortions of the cosmic microwave background}},  {\em JCAP} {\bf 1007}
  (2010) 019, [\href{http://xxx.lanl.gov/abs/0912.3655}{{\tt
  arXiv:0912.3655}}].

\bibitem{Stebbins:2007ve}
A.~Stebbins, {\it {CMB Spectral Distortions from the Scattering of Temperature
  Anisotropies}},  {\em Phys.Rev.D} (2007)
  [\href{http://xxx.lanl.gov/abs/astro-ph/0703541}{{\tt astro-ph/0703541}}].

\end{thebibliography}
\end{document}